\documentclass[aip,jcp,10pt,twocolumn]{revtex4-1}

\usepackage{times}
\usepackage{graphicx}
\usepackage{dcolumn}
\usepackage{amsmath}
\usepackage{bm}
\usepackage{amssymb}
\usepackage{subfig}
\usepackage{textcomp}
\usepackage{color}

\begin{document}

\title{Imaging state-to-state reactive scattering in the Ar$^+$ + H$_2$ charge transfer reaction} 



\author{Tim Michaelsen}

\affiliation{Institut f{\"u}r Ionenphysik und Angewandte Physik,
  Universit{\"a}t Innsbruck, Technikerstra{\ss}e 25/3, 6020
  Innsbruck, Austria}
\author{Bj\"orn Bastian}
\affiliation{Institut f{\"u}r Ionenphysik und Angewandte Physik,
  Universit{\"a}t Innsbruck, Technikerstra{\ss}e 25/3, 6020
  Innsbruck, Austria}
\author{Eduardo Carrascosa}
\affiliation{Institut f{\"u}r Ionenphysik und Angewandte Physik,
  Universit{\"a}t Innsbruck, Technikerstra{\ss}e 25/3, 6020
  Innsbruck, Austria}
\affiliation{The University of Melbourne, Parkville 3010 (VIC), Australia}
\author{Jennifer Meyer}
\affiliation{Institut f{\"u}r Ionenphysik und Angewandte Physik,
  Universit{\"a}t Innsbruck, Technikerstra{\ss}e 25/3, 6020
  Innsbruck, Austria}
\author{David H. Parker}
\affiliation{Department of Molecular and Laser Physics, Institute for Molecules and Materials, Radboud University, Heyendaalseweg 135, 6525 AJ Nijmegen, The Netherlands}
\author{Roland Wester}
\email{roland.wester@uibk.ac.at}
\affiliation{Institut f{\"u}r Ionenphysik und Angewandte Physik,
  Universit{\"a}t Innsbruck, Technikerstra{\ss}e 25/3, 6020
  Innsbruck, Austria}


\date{\today}

\begin{abstract}
The charge transfer reaction of Ar$^+$ with H$_2$ and D$_2$ has been investigated in an experiment combining crossed beams with three-dimensional velocity map imaging. Angle-differential cross sections for two collision energies have been obtained for both neutral species. We find that the product ions are highly internally excited. In the reaction with H$_2$ the spin-orbit excited Ar$^+$ state's coupling to the 'resonant' vibrationally excited product H$_2^+$($v=2$) dominates for both investigated energies, in line with previous investigations.  The observed angular distributions, however, show significantly less back-scattering than was found previously. Furthermore, we discovered that the product ions are highly rotationally excited. In the case of Ar$^+$ reacting with D$_2$ the energetically closest lying vibrational levels are not strictly preferred and higher-lying vibrational levels are also populated. For both species the backward-scattered products show higher internal excitation.
\end{abstract}


\maketitle 


\section{Introduction}

Ion-molecule reactions play an important role in many chemical environments such as interstellar molecular clouds and planetary atmospheres \cite{waite2007:sci,larsson2012:rpp}. Charge transfer processes in particular were found to be of importance for the hydrogen chemistry in the early universe \cite{Savin2004:a} and the X-ray emission of comets \cite{cravens2002:sci}.  To gain a mechanistic understanding of such reactions both theoretical and experimental studies of the involved dynamics are needed. To study charge transfer reaction dynamics we investigated one of the conceptually simplest systems: The reaction of Ar$^+$($^2$P$_J$) with H$_2$ and its isotope D$_2$. Argon is the third-most abundant component of the earth's atmosphere and is therefore an important reactant ion. Its reaction with molecular hydrogen has been a focus of many experimental and theoretical investigations over the last six decades and is among the most studied ion-molecule reactions \cite{Gutbier1957:zfuna, Stevenson1958:jcp, Giese1963:jcp, Klein1964:jcp, Aquilanti1965:jcp, Henglein1965:jcp, Lacmann1965:bdbfupc, Doverspike1966:jcp, Amme1966:jcp, Ding1967:bdbfupc, Fink1967:jcp, Mahadevan1968:pr, Hyatt1968:zn, Chupka1968:jcp, Bowers1969:jcp, Adams1970:jcp, Chiang1970:jcp, Ryan1973:jcp, Teloy1974:cp, Smith1976:ijmsip, HodgeJr1977:pra, Hierl1977:jcp, Lindinger1977:jcp, Rakshit1980:jcp, Tanaka1981:jcp, Dotan1982:jcp, Kemper1983:i, Kato1984:jcp, Hamdan1984:ijmsip, Ervin1985:jcp, Nakamura1986:jpsj, Henri1988:jcp, Lindsay1988:jpbamop, Liao1990:jcp, Gislason1991:jcp, Hawley1992:jcp, Tosi1993:jcp, Gerlich1993:acp, schweizer1994:ijm, Sizun1996:cp, Uiterwaal1996:cp, Sizun2002:jcp}:
\begin{align}
\rm{Ar^+(^2P_J) + H_2(X,v,J)} \nonumber \\ 
\longrightarrow ~ &\rm{ArH^+(X, v',J') + H(^2S_{1/2})} \\
\longrightarrow ~ &\rm{Ar(S_0) + H_2^+(X,v',J')} \\
\longrightarrow ~ &\rm{Ar^+(^2P_{J'}) + H_2(X,v',J')}
\label{reactions}
\end{align}
It can proceed via proton transfer, also referred to as atomic re-arrangement and indicated by equation (1), or via charge transfer as shown in equation (2). The charge transfer complex Ar-H$_2^+$ acts as an intermediate configuration for the proton transfer channel. The proton transfer and the charge transfer channel are exoergic by 1.5\,eV and 0.33\,eV, respectively. Ar$^+$ possesses two spin-orbit states, Ar$^+$($^2$P$_{3/2}$) and Ar$^+$($^2$P$_{1/2}$) that are separated by 178\,meV. Equation (3) describes a collision induced change in the Ar$^+$ spin-orbit state.

Early investigations of reaction (1) proposed a classical Langevin-type capture model\cite{Langevin1905:acp, Gioumousis1958:jcp, Stevenson1958:jcp}. Later temperature dependent studies revealed rates for reaction (1) smaller than the Langevin model predicts and a weak positive energy dependence over a large temperature range \cite{Adams1970:jcp, Dotan1982:jcp}. Spin-orbit state selective ionization techniques revealed an enhanced reactivity for spin-orbit excited Ar$^+$ compared to the ground state \cite{Tanaka1981:jcp, Henri1988:jcp, Liao1990:jcp, Uiterwaal1996:cp}. This enhancement is more pronounced for the charge transfer than for the proton transfer channel. The published ratios of the cross section of the spin-orbit excited compared to the ground state are in the range of 5 to 13 at collision energies below 1.5\,eV. For the proton transfer the ratios are 1.3 to 1.5 in the same energy range \cite{Liao1990:jcp}. At higher collision energies this enhancement diminishes and approaches 0.8 to 1.0 for both channels above 5\,eV.

The state-specific cross sections vary quite substantially for the mentioned experimental studies. Tanaka et al.\ obtained a cross section for the charge transfer of spin-orbit excited Ar$^+$ of 2.7\,\AA$^2$ at 0.48\,eV collision energy compared to the 16.2\,\AA$^2$ obtained by Liao et al.\ and 21.0\,\AA$^2$ obtained by Henri et al.\ at similar collision energies. It is interesting to note that the measured total cross sections for the combined reaction channels (1) and (2) exceed the Langevin cross section for the reaction of state selected Ar$^+$($^2$P$_{1/2}$) + H$_2$ at low collision energies in the experiments by Tanaka et al.\ and Liao et al.\,. Tanaka et al.\ explain this by arguing that the proton transfer proceeds first via a long-range electron jump followed by a reaction of H$_2^+$ with Argon. As Argon possesses a larger polarizability, the resulting Langevin rate is larger and fits as a limiting rate to their experimental results \cite{Tanaka1981:jcp}. The cross sections obtained by Ervin and Armentrout \cite{Ervin1985:jcp} as well as Tosi et al.\ \cite{Tosi1993:jcp} for the proton transfer channel with a statistical mixture of the two Ar$^+$ spin-orbit states are below the Langevin limit. The same is true for the the results obtained by Hawley et al.\ for reactions with state-selected ground state Ar$^+$ at lower collision energies \cite{Hawley1992:jcp}.

\begin{figure}
\includegraphics[width=0.9\columnwidth]{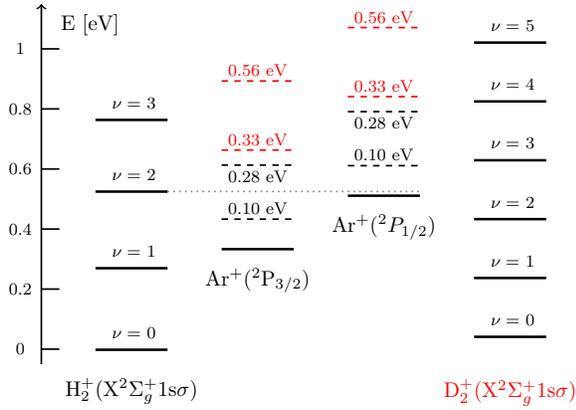}%
\caption{Schematic representation of the energy levels involved for both Ar$^+$ spin-orbit states and the vibrational levels of the H$_2^+$ and D$_2^+$ products. The investigated collision energies for reactions with H$_2$ are marked by black dashed lines and for reactions with D$_2$ as red dashed lines. The energy scale shown is given relative to the H$_2^+$ ground state.}
\label{levels}
\end{figure}

In addition to state selective ionization Liao et al.\ used the differential reactivity method to obtain the product ion vibrational state distribution. They found a clear preference for $v=2$ in reaction (2) for reactions with spin-orbit excited Ar$^+$. The enhanced reactivity and preference for $v=2$ in the charge transfer reaction (2) has been explained by a quasi resonance in energy. The energy of the educts Ar$^+$($^2$P$_{1/2}$) + H$_2$($X,v=0$) is very close to the energy of the vibrationally excited products Ar(S$_0$) + H$_2^+(v=2)$ ($\Delta E = 16$\,meV). The relevant energy levels are depicted in FIG.~\ref{levels}. This quasi-resonance in energy gives rise to a seam in the asymptotic part of the potential energy surfaces of the entrance and exit channel at intermolecular distances of approx.\ 4.4 to 5\,\AA \cite{Gislason1991:jcp, Uiterwaal1996:cp, Tosi1993:jcp}. This seam represents an avoided crossing through which the reaction of Ar$^+$($^2$P$_{1/2}$) can efficiently proceed to the vibrationally excited product H$_2^+(v=2)$. The decreasing contribution of this resonance effect at higher energies has been explained by the opening of other reaction pathways mediated by higher lying vibrational states.
In the conceptually similar charge transfer reaction of Ar$^+$ with N$_2$ the coupling to a vibrationally excited product state via an avoided crossing has been theoretically predicted and experimentally confirmed \cite{birkinshaw1987:cp, Liao1990:jcp, candori2001:jcp, Trippel2013:prl}. The preference for the H$_2^+(v=2)$ product is also evident in the reverse reaction as a significantly increased reaction rate if the H$_2^+$ is in the $v=2$ excited state \cite{Latimer1982:jpb, Houle1982:jcp, Kato1984:jcp, Liao1990:jcp}. The Ar$^+$ ground state ($^2$P$_{3/2}$) reacts slower than the spin-orbit excited state and couples most efficiently to H$_2^+(v=0)$ and $(v=1)$ at collision energies below 0.5\,eV \cite{Tanaka1981:jcp}.

In addition to reaction rate studies angle-differential cross sections have been published by Hierl et al.\ \cite{Hierl1977:jcp} They observed forward scattering for all investigated collision energies and an additional major backward contribution at a collision energy of 0.13\,eV. The forward-scattered product ions were observed close to the energy expected for the 'resonant' vibrational level of the H$_2^+$ products. A number of theoretical investigations have been carried out on the charge transfer and proton transfer reaction pathways \cite{Gislason1991:jcp, Tosi1993:jcp, Baer1990:jcp, Uiterwaal1996:cp,  Sizun1996:cp, Sizun2002:jcp}. Baer et al.\ \cite{Baer1990:jcp} published calculated reaction rates along with angle-differential cross sections based on a 'reactive infinite order sudden approximation (RIOSA)' approach. They predict forward and rainbow angle scattering for the charge transfer reaction. Recently, a quantum wave-packet calculation has been carried out by Hu et al. investigating the proton transfer channel of the reverse reaction H$_2^+$ + Ar \cite{Hu2013:jcp}. To our knowledge no state-of-the-art quantum scattering calculations have been published for reactions (1) and (2) and none of the mentioned theoretical work explicitly treated rotational excitation for these reaction pathways.

The isotopic reaction of Ar$^+$($^2$P$_J$) + D$_2$ has also been investigated in previous studies using various techniques \cite{Ervin1985:jcp, Tanaka1981:jcp, Hawley1992:jcp, schweizer1994:ijm}. The quasi-resonance is not present in this system and studies have shown that the reaction rate of the excited spin-orbit state for both reactions is only slightly enhanced compared to reactions with ground state Ar$^+$.

In this work we present angle- and energy-dependent differential cross sections of the charge transfer reaction obtained with a crossed-beam velocity map imaging spectrometer. It enables the detection of the full three-dimensional velocity distribution of the product ions in a kinematically complete way. From the observed velocities we extract the angular distribution of the reaction products and their excited rovibrational quantum states after the reaction.

\section{Method}

The experimental setup has been described elsewhere in more detail \cite{stei2016:natc} and only a brief description will be given here. The experiment combines the crossed-beam technique with a 3D-velocity map imaging (VMI) spectrometer to study ion-molecule reaction dynamics. It is operated at a repetition rate of 20\,Hz. The ions are produced either by electron impact in a supersonic expansion of Argon or by igniting a plasma in the expansion. Both techniques should produce a statistical distribution of Ar$^+$($^2$P$_{1/2}$) to Ar$^+$($^2$P$_{3/2}$) of 1:2. The ions are extracted, mass selected and guided into an octupole radio-frequency ion trap. Inside, through non-reactive collisions with a buffer gas (usually Argon), the kinetic energy spread of the ions is minimized and their spatial distribution confined using shaping electrodes. The ions are then extracted and decelerated before entering the interaction region where they cross a molecular beam, produced by a supersonic expansion, under an angle of 60\textdegree. The molecular beam is produced by a pulsed piezo-cantilever valve and skimmed before crossing the ion-packet. The overlap region is located in the center of a VMI electrode stack which is pulsed on to extract the product ions perpendicular to the interaction plane. The velocity in the scattering plane is mapped onto a position-sensitive detector consisting of two multi-channel-plates and a phosphor screen. The impact-position of the ions is detected by a CCD-camera. Additionally, we record the flight-time by measuring the arrival of the photons emitted from the phosphor-screen with a photo-multiplier-tube. The raw laboratory images are transformed into a center-of-mass velocity distribution. Together with the time-of-flight information the full three-dimensional velocity vector for each event can be extracted. The maximum velocity is given by the available energy, namely the exothermicity plus the collision energy, and is referred to as the kinematic cut off. If the measured velocity is smaller than this value, due to conservation of energy, we can attribute the difference to energy transfer into internal excitation of the molecular products. Although we obtain the full three-dimensional velocity distribution we usually represent them as a 2D distribution by projecting all ion velocity vectors into a plane given by the  velocity along the center-of-mass axis v$_{\text{x}}$ and a radial velocity v$_{\text{r}}$. In this center-of-mass frame the neutral beam travels to the left.

The velocity and energy spread of the reactant ion beam are also obtained with the VMI technique. Typical values for the energy spread (FWHM) of the Ar$^+$ beam are in the order of 160-250\,meV. In order to obtain the corresponding quantities for the neutral beam, it is ionized by an electron gun prior to detection with the VMI spectrometer. For the normal-H$_2$ neutral beam the spread is around 40\,meV corresponding to a translational temperature of approx.\ 70\,K. The obtained velocity distributions are broadened by the energy spread of the two reactant beams along with smearing caused by imperfect imaging of the ions. We can estimate this broadening from the experimental parameters obtained for the two beams and simulation results of the VMI conditions\cite{stei2013:jcp, Wester2014:pccp}. Under the experimental conditions of the presented experiments the energy spread of the beams dominates this broadening and amounts to below 20\,meV (FWHM) for forward-scattered ions at the observed velocities.

\section{Results}

\begin{figure}
 \centering
 \subfloat{%
   \includegraphics[width=0.5\columnwidth]{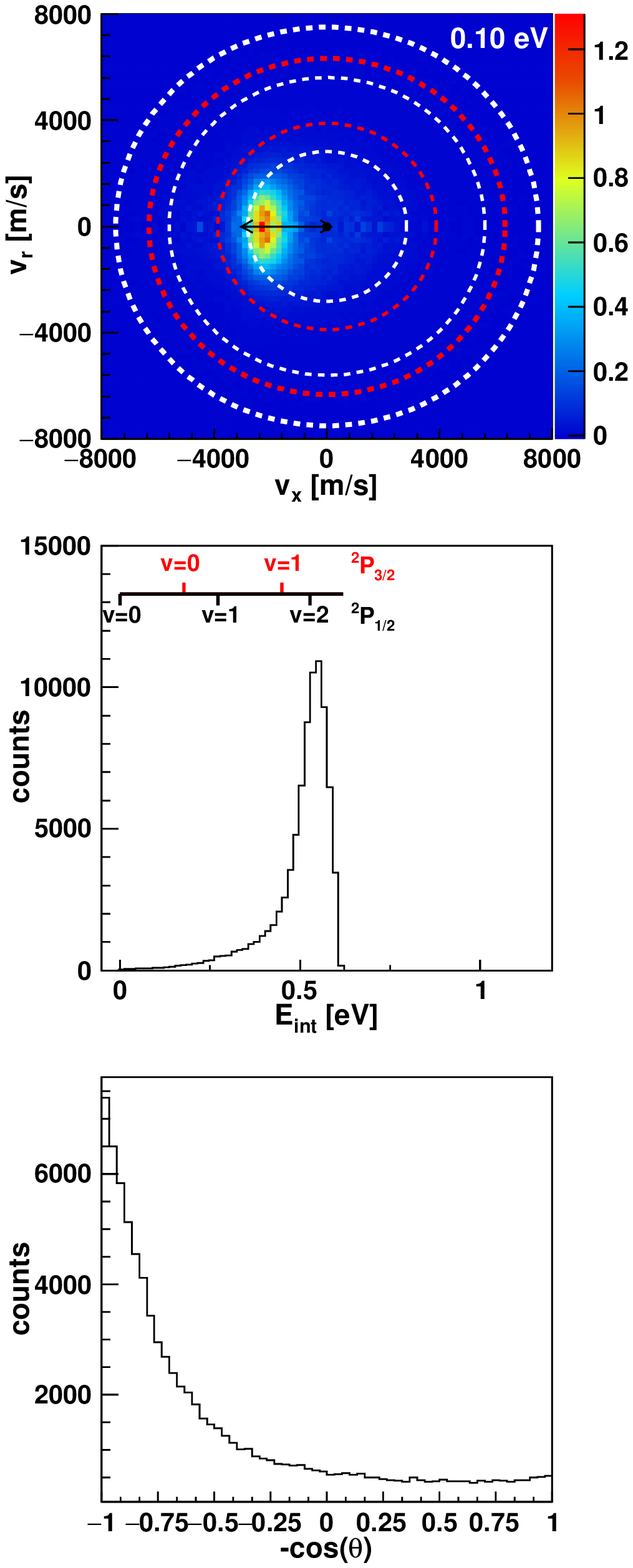}
 }
 \subfloat{%
   \includegraphics[width=0.5\columnwidth]{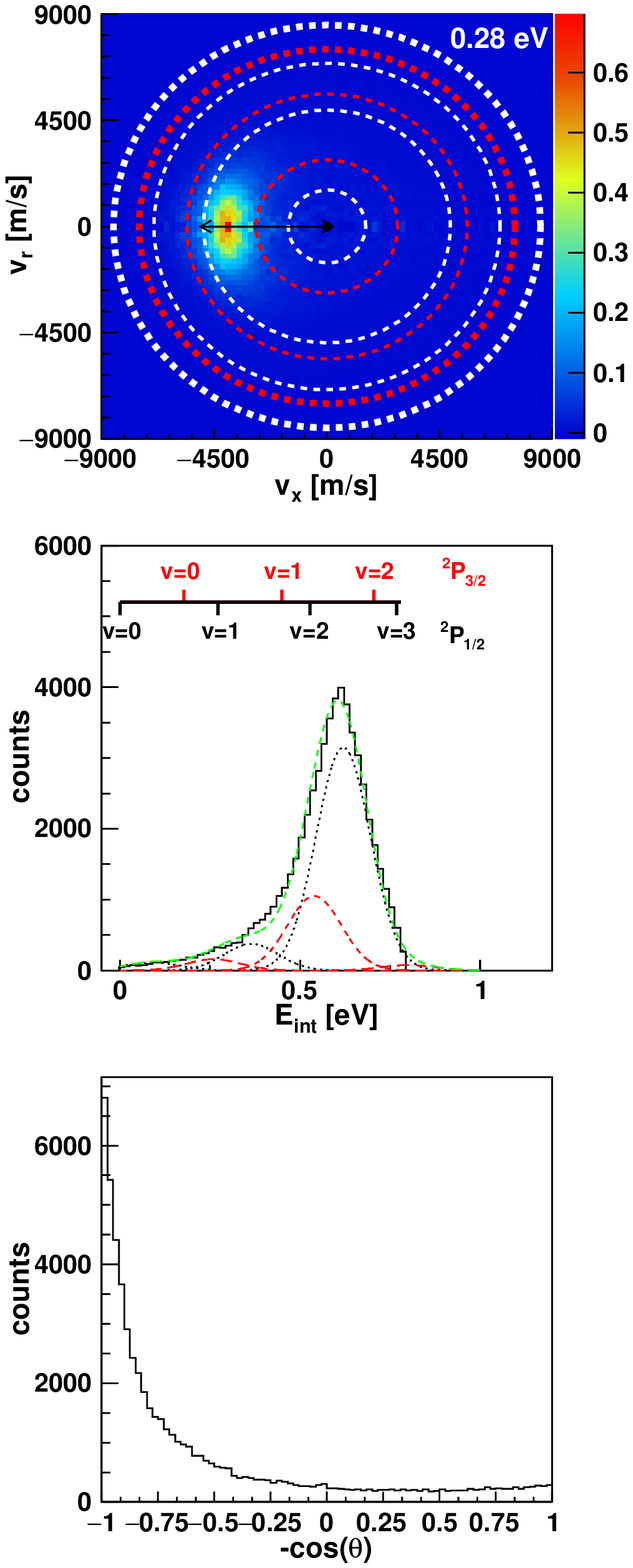}
 }
 \caption{H$_2^+$ velocity distribution, internal energy and angular distribution at 0.1\,eV (left) and 0.28\,eV (right) relative collision energy. The rings drawn into the velocity distribution  correspond to the kinematic cutoff for Ar$^+$($^2$P$_{1/2}$) (white rings) and Ar$^+$($^2$P$_{3/2}$) (red rings) and the corresponding vibrational levels of H$_2^+$. The positions of these levels are also shown as an inset axis in the internal energy distribution. The internal energy distribution for 0.28\,eV collision energy also shows the fit of a sum of Gaussians to the measured distribution (see text for details).}
 \label{fig:H2}
\end{figure}

The results of the charge transfer reaction of Ar$^+$($^2$P$_{3/2,1/2}$) + H$_2$ are presented in FIG.~\ref{fig:H2} alongside the extracted internal energy distribution of the H$_2^+$ products and the angular distribution represented by -cos($\theta$) to be consistent with the neutral beam direction for 0.1\,eV and 0.28\,eV collision energies. The outermost ring drawn into the velocity distribution corresponds to the kinematic cut off and the inner rings represent the vibrational levels of the H$_2^+$ products (see also FIG.~\ref{levels}). The white rings correspond to the spin-orbit excited Ar$^+$($^2$P$_{1/2}$) and the red rings to Ar$^+$($^2$P$_{3/2}$). The arrows mark the mean velocity of the neutral and ion beam prior to collision in the center-of-mass frame. In the second panel we present the internal energy distribution along with an inset axis marking the vibrational levels of H$_2^+$ for reactions with both spin-orbit states. It is presented, here and in all of the following plots, relative to the kinematic cutoff of Ar$^+$($^2$P$_{1/2}$). We observe predominantly forward scattering for both energies with a small backwards fraction. The internal energy distribution is sharply peaked corresponding to an internal excitation that lies slightly above the energy expected for the excitation of $v=2$ resulting from a reaction with Ar$^+$($^2$P$_{1/2}$) for both collision energies. We expect rotational excitation of the H$_2^+$ resulting in additional internal excitation, which explains the shift of the peak to higher internal energy. The width of the peak at 0.1\,eV relative energy is 80\,meV (FWHM) and therefore several times broader than the experimental broadening. This lets us conclude that multiple states, which are individually broadened by the experimental uncertainties, contribute to the total width. At 0.28\,eV collision energy the internal energy distribution is broader (182\,meV), slightly shifted to higher internal energy and more symmetric. At this collision energy two more state combinations are energetically allowed: H$_2^+$($v=2$) resulting from a reaction with Ar$^+$($^2$P$_{3/2}$) at an internal energy of 0.71\,eV and H$_2^+$($v=3$) after reaction with Ar$^+$($^2$P$_{1/2}$) at 0.77\,eV.  

From the internal energy distribution, we can conclude that a lot of energy is transferred to internal excitation. The charge transfer reaction is exothermic by 0.51\,eV regarding the spin-orbit excited state. All of the released energy is converted into internal excitation and some additional energy is converted from initial kinetic energy to internal degrees of freedom. This leads to slower average velocities of the H$_2^+$ products compared to the initial neutral molecules. This difference can be clearly seen when we compare the initial mean velocity of the neutral beam, represented by the black arrow pointing left drawn into the velocity distribution in the top panel of FIG.~\ref{fig:H2}, to the position of the main peak of the H$_2^+$ product distribution. 

In the low energy part of the internal energy distribution other state combinations contribute. Reactions with Ar$^+$($^2$P$_{3/2}$) resulting in H$_2^+$($v=1$) should appear at 0.45\,eV and with Ar$^+$($^2$P$_{1/2}$) resulting in  H$_2^+$($v=1$) at 0.28\,eV. The high energy part of the internal energy is limited by the maximum available energy: 0.61\,eV and 0.79\,eV respectively. This explains the sharp drop and the resulting asymmetric peak shape in the 0.1\,eV collision energy data.

\begin{figure}
 \includegraphics[width=1\columnwidth]{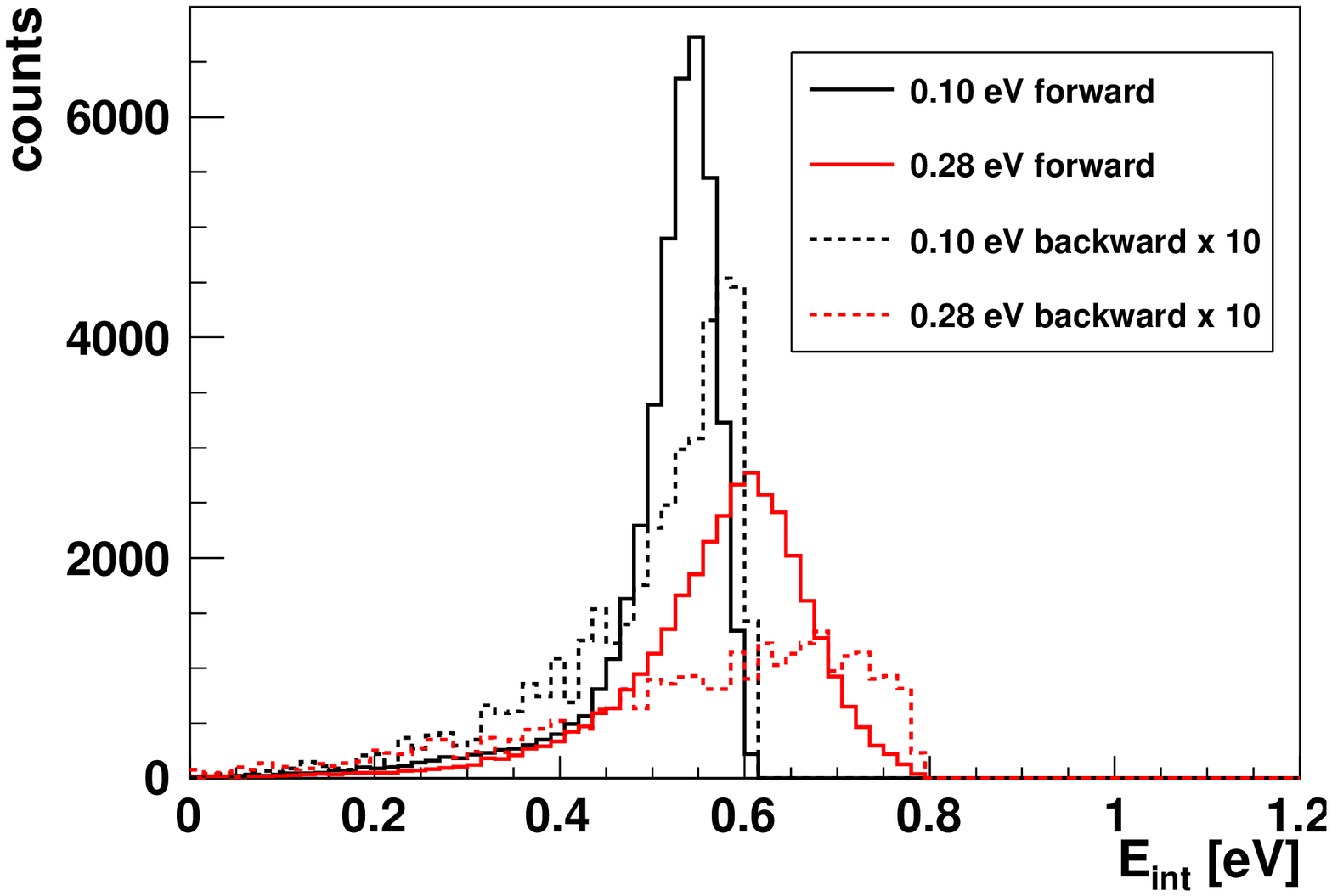} \\
 \includegraphics[width=1\columnwidth]{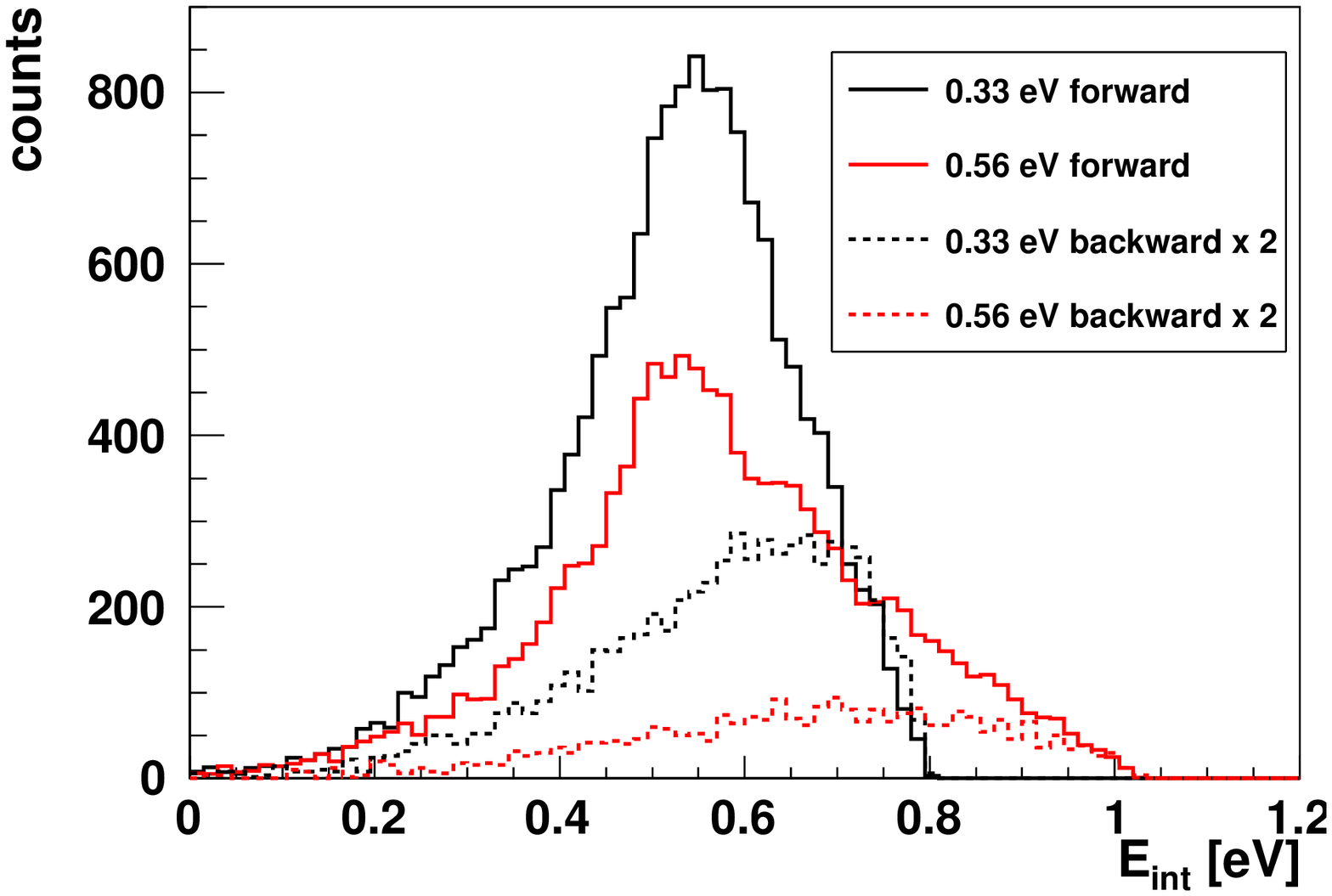}%
 \caption{Internal energy distributions of H$_2^+$ (top) and D$_2^+$ (bottom) from the charge transfer reaction with Ar$^+$ for a 45\textdegree\ cone in forward direction (solid lines) and backward direction (dashed lines). The backward contribution is enhanced by a factor of 10 in the top panel and by a factor of 2 in the bottom panel.}%
 \label{Eint_backward}
\end{figure}

From the angular distributions we can immediately see that the backwards contribution is very small compared to the forward-scattered peak. At the lower collision energy a larger backward-scattered fraction as well as scattering into larger angles is observed. The backward-scattered peak lies at even lower velocities corresponding to higher internal excitation of the H$_2^+$ ions. As this cannot be easily inferred from the scattering image due to the low fraction of backward-scattered product ions, we show the internal energy distribution for a 45\textdegree\ cone in forward (solid lines) and backward (dashed lines) direction in FIG.~\ref{Eint_backward}. The backward fraction is enhanced by a factor of ten. One can clearly see a shift towards higher internal energy in the backward-scattered fraction.
 
\begin{figure}
 \centering
 \subfloat{%
   \includegraphics[width=0.5\columnwidth]{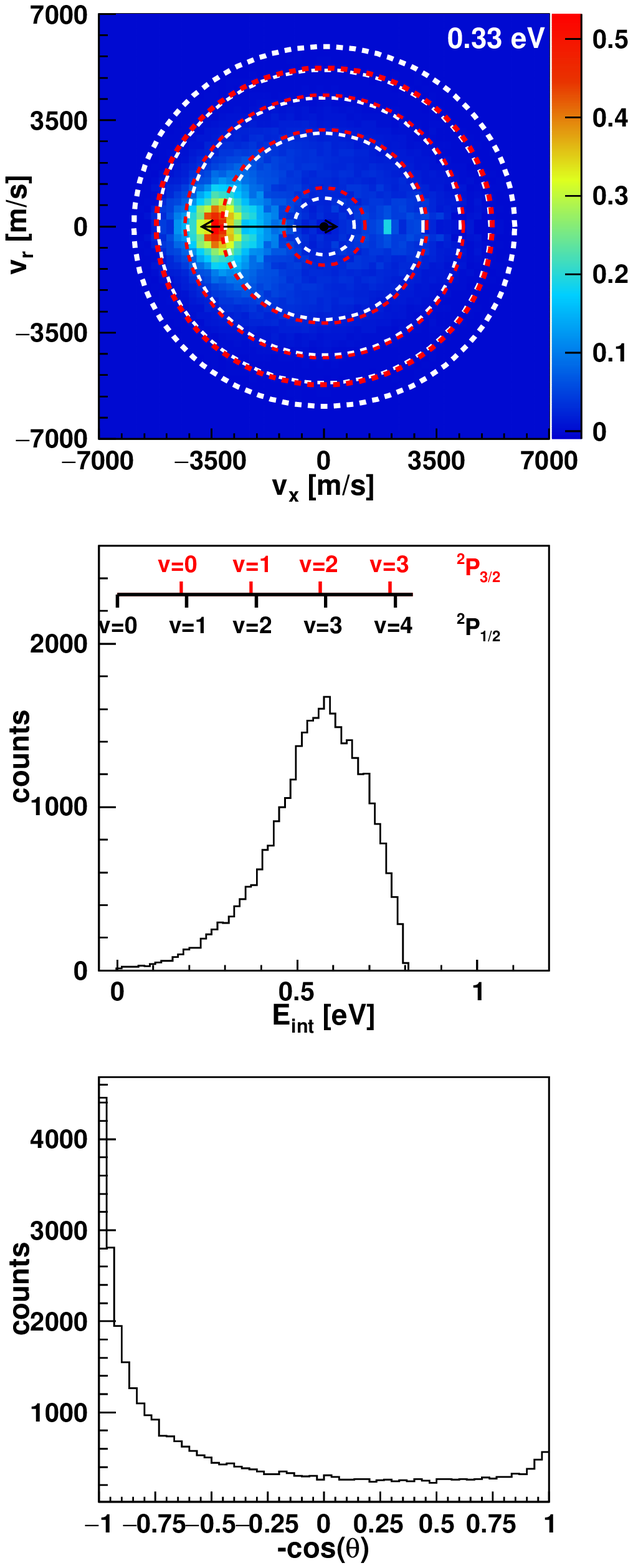}
 }
 \subfloat{%
   \includegraphics[width=0.5\columnwidth]{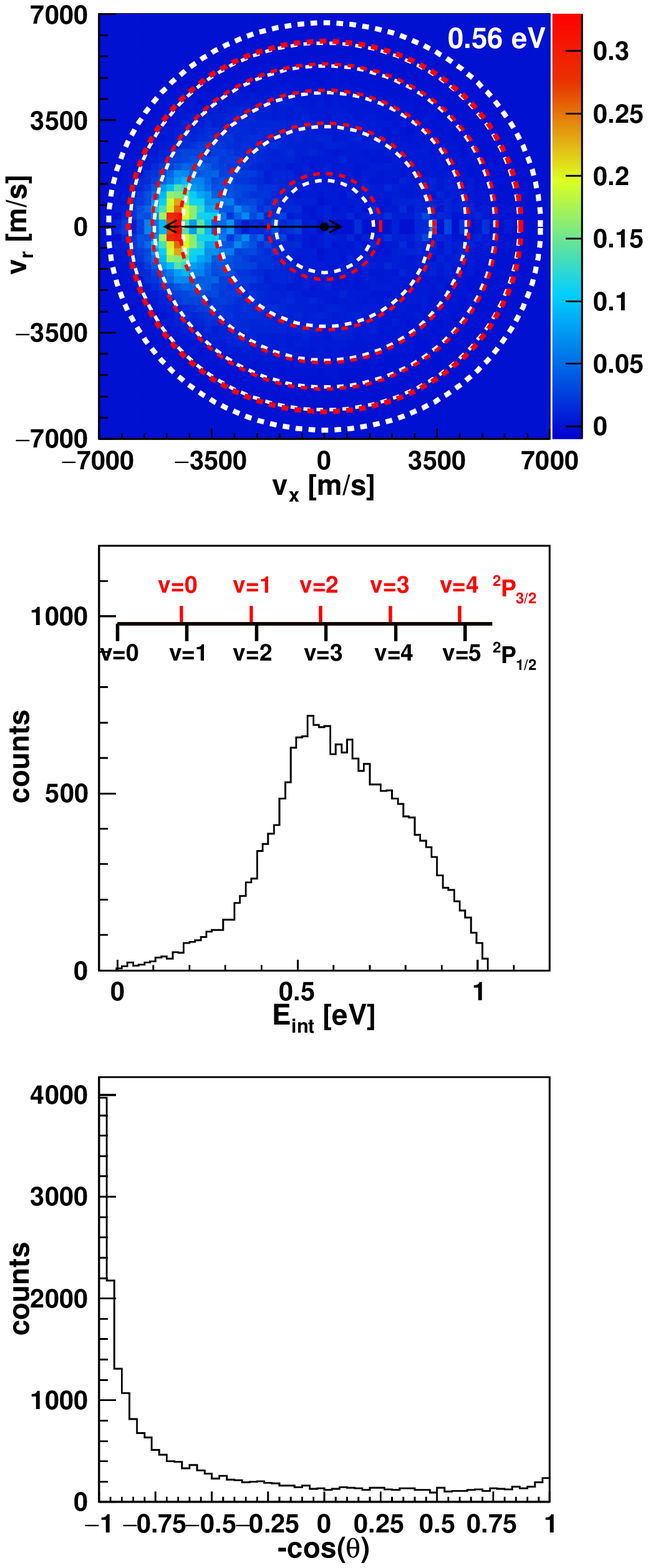}
 }
 \caption{D$_2^+$ velocity distribution, internal energy and angular distribution at 0.33\,eV (left) and 0.56\,eV (right) relative energy. The rings drawn into the velocity distribution correspond to the kinematic cutoff for Ar$^+$($^2$P$_{1/2}$) (white rings) and Ar$^+$($^2$P$_{3/2}$) (red rings) and the corresponding vibrational levels of D$_2^+$. The positions of these levels are also shown as an inset axis in the internal energy distribution.}
 \label{fig:D2}
\end{figure}

The results for the charge transfer reaction with D$_2$ are depicted in FIG. \ref{fig:D2}. The rings drawn into the center-of-mass velocity distribution again correspond to the kinematic cutoff and the vibrational levels of D$_2^+$ for reactions with Ar$^+$($^2$P$_{1/2}$) (white) and  Ar$^+$($^2$P$_{3/2}$) (red). The vibrational spacing in D$_2^+$ is very close to the energy spacing between the two spin-orbit states of Ar$^+$. This makes it challening to disentangle the initial Ar$^+$ state that is involved in the reaction. 

We observe forward scattering with a small backward contribution. At the lower investigated collision energy of 0.33\,eV the forward peak position lies energetically above the internal energy corresponding to the D$_2^+$ vibrational level $v=1$ and $v=2$ for a reaction with Ar$^+$($^2$P$_{3/2}$) and Ar$^+$($^2$P$_{1/2}$) respectively. The two higher-lying vibrational levels $v=2$ and $v=3$ also overlap with the distribution. At 0.56\,eV relative energy the peak is a lot broader and extends to higher internal energy. At this collision energy higher-lying vibrational states are energetically accessible and contribute to the internal energy distribution. The closest vibrational levels of D$_2^+$, energetically speaking, are $v=1$ for Ar$^+$($^2$P$_{3/2}$) and $v=2$ for Ar$^+$($^2$P$_{1/2}$) (see FIG. \ref{levels}). While it appears that they contribute, especially considering rotational excitation, higher vibrational levels are significantly populated. At the higher collision energy this effect is especially evident. In this regard the data obtained for the reaction with D$_2$ differs from the experiments with H$_2$, where the quasi-resonant vibrational state is predominantly excited.

As in the reaction with H$_2$ the backward fraction is larger at lower collision energy. Overall the back-scattered fraction is larger in the reaction with D$_2$ compared to the H$_2$ results. The internal energy distributions are significantly broader compared to the reaction with H$_2$. This broadening partially stems from the fact that the presented internal energy plots are integrated over all angles. One can already infer from the scattering image that the backward-scattered ions appear at lower velocities. Thus, they are even higher internally excited and appear at higher energies in the E$_{\text{int}}$ plot. This behavior is more pronounced in the D$_2$ data due to the larger backward-scattered fraction, but is also present for the reaction with H$_2$. To illustrate this we present the internal energy distribution of a 45\textdegree\ cone cut from the total differential cross sections in both forward and backward direction in the lower panel of FIG.~\ref{Eint_backward}. The backward contribution is enhanced by a factor of 2 in this plot. We observe a clear shift towards higher energy in the back-scattered fraction. At the lower collision energy the purely forward-scattered peak is much more symmetric compared to the peak integrated over all angles shown in FIG.~\ref{fig:D2}. At 0.56\,eV collision energy the forward-scattered peak does not appear more symmetric, instead the right slope appears almost bimodal and the decrease is less steep, when compared with the lower energy and the H$_2$ data. This might be explained by larger contributions of the $v=3$ and $v=4$ as well as $v=4$ and $v=5$ vibrational levels for Ar$^+$($^2$P$_{3/2}$) and Ar$^+$($^2$P$_{1/2}$), respectively.

\section{Discussion}

The majority of the previous studies on this reaction focused on obtaining total or state-specific cross sections. In order to compare these to our results we extract the approximate amount of product ions in specific vibrational states for both spin orbit states of the reacting Ar$^+$ from the internal energy distribution of the product ions. To this end we fit seven Gaussian functions, one for each combination of quantum states, to the measured energy distribution at 0.28\,eV collision energy (see FIG.~\ref{fig:H2}, right column, central panel). Each Gaussian has a width of 150\,meV(FWHM) and is positioned to the right of the respective quantum state marker to account for rotational excitation (see below). Assuming a statistical 2:1 ratio of Ar$^+$($^2$P$_{3/2}$) to Ar$^+$($^2$P$_{1/2}$) reactants, we obtain from the fit that about 85\% of the product ions originate from reactions with Ar$^+$($^2$P$_{1/2}$). More than 90\% of these product ions are excited to $v=2$. Correspondingly, we estimate the total contribution from reactions with Ar$^+$($^2$P$_{3/2}$) to be approx.\ 15\%. Here the majority ($\sim$ 90\%) is excited to $v=1$. For the cross section ratio $\sigma_{P_{1/2}}$ / $\sigma_{P_{3/2}}$ this yields a value of about 6. This is in good agreement with previous studies, which found ratios of 5 to 13 at similar collision energies \cite{Liao1990:jcp}.

We also compare our results with the one previous angle- and energy differential cross section measurement of the charge transfer channel. Hierl et al.\ investigated the reaction at three relative collision energies 0.13, 0.48 and 3.44\,eV using rotatable crossed beams and a fixed detector arrangement \cite{Hierl1977:jcp}. They used a thermal hydrogen beam emitted by a multi capillary array as their neutral beam source. From their published angular distribution at 0.13\,eV we estimate the backward-scattered fraction, compared to the amount of forward-scattered product ions, to be larger than 70\%. Our experiments show a significantly smaller backward-scattered fraction below 10\% at a collision energy of 0.1\,eV. While a similar trend of decreasing backward-scattered fraction with increasing collision energy is observed in both experiments, there is also a disagreement for the higher energy data. In their data for 0.48\,eV collision energy a higher fraction ($>$ 15\%) is scattered backwards compared to our data for 0.3\,eV (8\%). In their published velocity contour maps the position of the main peak in forward direction lies slightly above the velocity corresponding to H$_2^+$($v=2$), which they label as 'resonant charge transfer (RCT)'. In our measurements the main peak lies below this velocity. We attribute this additional internal excitation in our results to rotational excitation as discussed below. The shift in energy might be due to the thermal H$_2$ beam used in their experiments, when compared to our data obtained from collisions with a supersonic, colder H$_2$ beam.

Both Hierl et al.\ and the present work observe scattering into larger angles and enhanced backward scattering at lower collision energies. This indicates that small impact parameter collisions play a more important role at lower collision energies. A similar attribution of scattering into large angles to small impact parameter collisions was also done by Trippel et al.\ in the investigation of the Ar$^+$ + N$_2$ charge transfer reaction \cite{Trippel2013:prl}. The main peak in the backward-scattered fraction is observed at lower product velocities than the forward peak, corresponding to even more transfer of kinetic energy into internal excitation, as already discussed in the results section. This shift for the peak in backward direction has also been observed in the experiments by Hierl et al.\ .
 
The simulated angular distribution published by Liao et al.\ shows forward scattering with the addition of rainbow-angle scattering at relatively high collision energies (1.44\,eV) \cite{Liao1990:jcp}. This feature would smear out due to the limited resolution of our experiment and the presence of rotational excitation, but should still appear as a 'dip' in the angular distribution. We do not observe any change in the slope of the angular distributions at angles between 25-30\textdegree\ and therefore do not observe any indication of rainbow scattering.

In the following we consider the role of product rotational excitation in the studied reaction. As noted above, we attribute the shift of the forward-scattered peak from the expected quasi-resonant vibrational level to additional rotational excitation. To determine the amount of excitation we have to assess the rotational state population prior to reaction.  Previous studies on rotational cooling of H$_2$ in supersonic expansions have shown that the rotational state population cools almost completely to the ortho- and para-H$_2$ ground states \cite{Ramos2009:jpca}. Most of the population was found to end up in $J=0$ and $J=1$ for para- and ortho-H$_2$ respectively with contributions of a few percent in $J=2$ and $J=3$. Based on this we expect a statistical mixture of mostly ground state ortho- and para-H$_2$, specifically in a ratio of 3:1 and the two species can not inter-convert through reactive collisions. 

\begin{figure}
 \includegraphics[width=1\columnwidth]{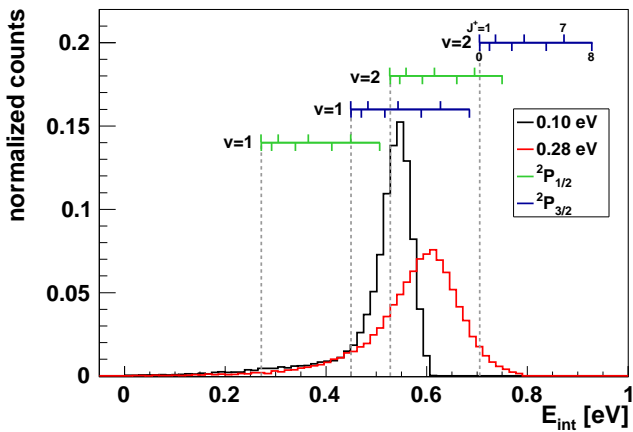}%
 \caption{Internal energy distributions of the charge transfer reaction with H$_2$. The inset axes mark rotational levels of the H$_2^+$ products for the involved Ar$^+$ spin-orbit states and H$_2^+$ vibrational levels. Para-H$_2$ rotational levels are depicted as downward-pointing and ortho-H$_2$ levels as upward-pointing ticks.}%
 \label{Eint_rotations}
\end{figure}

We present the area-normalized internal energy distributions of the reaction with H$_2$ in FIG.~\ref{Eint_rotations}. To exclude contributions from the backward fraction we only consider product ions in a forward cone with a half-angle of 30\textdegree. When analyzing the product rotational states we assume that the rotational angular momentum quantum number can only change by a multiple of two due to the symmetry of the interaction potential. The four inset axes mark the energy of rotational states $J^+$ of H$_2^+$ corresponding to different originating Ar$^+$ spin-orbit and vibrational states up to $J^+=8$. The upward-pointing ticks show the odd-numbered rotational levels for collisions with ortho-H$_2$ and the downward-pointing ticks show the energy of the rotational levels when colliding with para-H$_2$. The vertical dotted lines mark the rotational ground state of H$_2^+$ ($J^+=0$) and serve only as a visual guide. In the following discussion we often refer to product rovibrational states ($v,J^+$) along with the originating spin-orbit state Ar$^+$($^2$P$_\text{J}$)  and will use the notation ($v,J^+$)$_{\text{P$_\text{J}$}}$. Spin-rotation coupling in the hydrogen ions is very small and safely ignored. The peak maximum at 0.1\,eV collision energy corresponds to an energy close to ($v=2,J^+=2$)$_{\text{P$_{1/2}$}}$. We interpret the peak as an overlap of population in $J^+=1-4$ of this vibrational level as well as contributions from rotational levels of the vibrational level ($v=1$)$_{\text{P$_{3/2}$}}$. The maximum available energy is very close to ($v=2,J^+=5$)$_{P_{1/2}}$. Despite the lower reaction rate a contribution from ($v=1,J^+$)$_{P_{3/2}}$ is not surprising, because there is twice as much population in this spin-orbit state. We attribute the slowly increasing slope in the low internal energy part to contributions from ($v=1,J^+$)$_{P_{1/2}}$. Only highly excited rotational levels of this state overlap with the main peak.

In the internal energy distribution corresponding to the higher relative energy of 0.28\,eV shown in red, the peak is broader and is shifted towards higher internal excitation. The center of the peak lies close to ($v=2,J^+=4$)$_{P_{1/2}}$ and  ($v=2,J^+=5$)$_{P_{1/2}}$. The maximum available energy of 0.78\,eV allows for rotational states of this vibrational level up to $J^+=8$ to be populated and contributions from another state in the right slope: ($v=2,J^+$)$_{P_{3/2}}$. We can conclude that a large fraction of the additional translational energy is converted into internal energy and that even higher rotational states are excited. Given that most of the neutral H$_2$ molecules are in the lowest ortho- and para-states, efficient excitation of multiple quanta of rotation in one reactive collision is needed to explain the observed internal energy distributions.

To estimate the orbital angular momentum of the reactants we consider the charge transfer distance of about 5\,\AA\ as an upper limit for the impact parameter of the Ar$^+$ + H$_2$ collision. As described in the introduction this value is larger than the maximum impact parameter obtained from the Langevin model. At the collision energy of 0.28\,eV we estimate the angular momentum for this maximum impact parameter to be about 90\,$\hbar$. For the forward-scattered flux we assume a minimum impact parameter of at least 1.5\,\AA, based on a simple geometric estimate of the extent of the repulsive wall of the interaction potential. Even at this lower limit the total angular momentum is about 25\,$\hbar$. Thus, the angular momentum available to forward scattering is clearly large enough to allow for the rotational excitation up to $J=8$ for H$_2^+$ in $v=2$ at 0.28\,eV collision energy. Higher rotational excitations would also be possible given the available angular momentum, but they are energetically inaccessible at this vibrational level.

As a model to rationalize the large angular momentum transfer from orbital angular momentum to hydrogen rotation we suggest a two-step process. At the large internuclear separation around the charge transfer distance an electron moves from H$_2$ to Ar$^+$. This changes the interaction potential to an effective Ar-H$_2^+$ potential, which is much more attractive than the Ar$^+$-H$_2$ potential as mentioned in the introduction. Rotationally inelastic scattering on this potential energy surface may then lead to the rotational excitation of the hydrogen molecular ion. Such inelastic cross sections can be substantial, as recently calculated for the He-H$_2^+$ system \cite{hernandez2017:jcp}.

For backward-scattered products we observe even more energy partitioning into internal excitation. At an impact parameter of 0.5\,\AA \,the angular momentum is still 8\,$\hbar$ meaning an excitation to high rotational levels is still feasible but must be vastly more efficient than for the forward-cattered products. In line with the suggested two-step model this implies that this efficiency, which depends on the details of the involved potential surface, is substantially different in the repulsive region at short range. Detailed theoretical calculations are needed to test this model quantitatively in comparison with the experimental data.

\section{Conclusion}

We obtained angle- and energy-differential cross sections for the charge transfer reaction of Ar$^+$($^2$P$_J$) with H$_2$(X$,v=0,J$) and D$_2$(X$,v=0,J$). In both reactions we observe predominantly forward scattering with a small backward-scattered fraction. In the case of reactions with H$_2$ we can attribute most of the H$_2^+$ product ions to reactions with spin-orbit excited Ar$^+$ resulting in vibrationally excited H$_2^+$ in $v=2$ for both investigated collision energies. This resonant coupling into a specific state is likely due to the small energy difference between the spin-orbit excited state and this vibrational state of the product.

For the isotopic reaction no energetically resonant state exists as the vibrational states of D$_2$ are shifted compared to H$_2$. This changes the observed product state distribution. Instead of predominant coupling to a single product state we observe excitation of multiple vibrational states.

We also find that the product ions are rotationally excited. At the lower relative energy studied in the reaction with H$_2$ they are excited up to the maximum available rotational state. At the higher collision energy multiple rotational quanta are excited in a single reactive collision. We observe a large transfer of energy into internal excitation, even of initial kinetic energy, despite the exoergic nature of the reaction. At low relative energy we observe scattering into larger angles and an increased amount of back-scattered products for both neutral species. The products that are scattered backwards show even more internal excitation. For the reaction with D$_2$ we observe an overall increase in the backward-scattered fraction when compared to H$_2$.

We have suggested a simple two-step model to rationalize the rotational excitation in the molecular reaction product. Detailed quantum scattering calculations would be helpful to test this. Furthermore, they could help understand the discrepancy in the cross section for backward and forward-scattered products and the disparity in the internal excitation of the forward and backward flux. It seems despite the decade-spanning study of this fundamental three-atom ion-molecule reaction we still do not fully understand the mechanisms involved and we hope this work will stimulate further investigations.


%
%

%

\begin{acknowledgments}
This work is supported by the Austrian Science Fund (FWF), project P25956-N20. Eduardo Carrascosa acknowledges funding by the DOC-scholarship of the Austrian Academy of Sciences (OeAW).
\end{acknowledgments}


%


\begin{thebibliography}{60}%
\makeatletter
\providecommand \@ifxundefined [1]{%
 \@ifx{#1\undefined}
}%
\providecommand \@ifnum [1]{%
 \ifnum #1\expandafter \@firstoftwo
 \else \expandafter \@secondoftwo
 \fi
}%
\providecommand \@ifx [1]{%
 \ifx #1\expandafter \@firstoftwo
 \else \expandafter \@secondoftwo
 \fi
}%
\providecommand \natexlab [1]{#1}%
\providecommand \enquote  [1]{``#1''}%
\providecommand \bibnamefont  [1]{#1}%
\providecommand \bibfnamefont [1]{#1}%
\providecommand \citenamefont [1]{#1}%
\providecommand \href@noop [0]{\@secondoftwo}%
\providecommand \href [0]{\begingroup \@sanitize@url \@href}%
\providecommand \@href[1]{\@@startlink{#1}\@@href}%
\providecommand \@@href[1]{\endgroup#1\@@endlink}%
\providecommand \@sanitize@url [0]{\catcode `\\12\catcode `\$12\catcode
  `\&12\catcode `\#12\catcode `\^12\catcode `\_12\catcode `\%12\relax}%
\providecommand \@@startlink[1]{}%
\providecommand \@@endlink[0]{}%
\providecommand \url  [0]{\begingroup\@sanitize@url \@url }%
\providecommand \@url [1]{\endgroup\@href {#1}{\urlprefix }}%
\providecommand \urlprefix  [0]{URL }%
\providecommand \Eprint [0]{\href }%
\providecommand \doibase [0]{http://dx.doi.org/}%
\providecommand \selectlanguage [0]{\@gobble}%
\providecommand \bibinfo  [0]{\@secondoftwo}%
\providecommand \bibfield  [0]{\@secondoftwo}%
\providecommand \translation [1]{[#1]}%
\providecommand \BibitemOpen [0]{}%
\providecommand \bibitemStop [0]{}%
\providecommand \bibitemNoStop [0]{.\EOS\space}%
\providecommand \EOS [0]{\spacefactor3000\relax}%
\providecommand \BibitemShut  [1]{\csname bibitem#1\endcsname}%
\let\auto@bib@innerbib\@empty
\bibitem [{\citenamefont {J.~H.~Waite}\ \emph {et~al.}(2007)\citenamefont
  {J.~H.~Waite}, \citenamefont {Young}, \citenamefont {Cravens}, \citenamefont
  {Coates}, \citenamefont {Crary}, \citenamefont {Magee},\ and\ \citenamefont
  {Westlake}}]{waite2007:sci}%
  \BibitemOpen
  \bibfield  {author} {\bibinfo {author} {\bibfnamefont {J.}~\bibnamefont
  {J.~H.~Waite}}, \bibinfo {author} {\bibfnamefont {D.~T.}\ \bibnamefont
  {Young}}, \bibinfo {author} {\bibfnamefont {T.~E.}\ \bibnamefont {Cravens}},
  \bibinfo {author} {\bibfnamefont {A.~J.}\ \bibnamefont {Coates}}, \bibinfo
  {author} {\bibfnamefont {F.~J.}\ \bibnamefont {Crary}}, \bibinfo {author}
  {\bibfnamefont {B.}~\bibnamefont {Magee}}, \ and\ \bibinfo {author}
  {\bibfnamefont {J.}~\bibnamefont {Westlake}},\ }\href {\doibase
  10.1126/science.1139727} {\bibfield  {journal} {\bibinfo  {journal}
  {Science}\ }\textbf {\bibinfo {volume} {316}},\ \bibinfo {pages} {870}
  (\bibinfo {year} {2007})}\BibitemShut {NoStop}%
\bibitem [{\citenamefont {Larsson}, \citenamefont {Geppert},\ and\
  \citenamefont {Nyman}(2012)}]{larsson2012:rpp}%
  \BibitemOpen
  \bibfield  {author} {\bibinfo {author} {\bibfnamefont {M.}~\bibnamefont
  {Larsson}}, \bibinfo {author} {\bibfnamefont {W.}~\bibnamefont {Geppert}}, \
  and\ \bibinfo {author} {\bibfnamefont {G.}~\bibnamefont {Nyman}},\ }\href
  {\doibase /10.1088/0034-4885/75/6/066901} {\bibfield  {journal} {\bibinfo
  {journal} {Rep. Prog. Phys.}\ }\textbf {\bibinfo {volume} {75}},\ \bibinfo
  {pages} {066901} (\bibinfo {year} {2012})}\BibitemShut {NoStop}%
\bibitem [{\citenamefont {Savin}\ \emph {et~al.}(2004)\citenamefont {Savin},
  \citenamefont {Krsti{\'c}}, \citenamefont {Haiman},\ and\ \citenamefont
  {Stancil}}]{Savin2004:a}%
  \BibitemOpen
  \bibfield  {author} {\bibinfo {author} {\bibfnamefont {D.~W.}\ \bibnamefont
  {Savin}}, \bibinfo {author} {\bibfnamefont {P.~S.}\ \bibnamefont
  {Krsti{\'c}}}, \bibinfo {author} {\bibfnamefont {Z.}~\bibnamefont {Haiman}},
  \ and\ \bibinfo {author} {\bibfnamefont {P.~C.}\ \bibnamefont {Stancil}},\
  }\href {\doibase /10.1086/421108} {\bibfield  {journal} {\bibinfo  {journal}
  {ApJL}\ }\textbf {\bibinfo {volume} {606}},\ \bibinfo {pages} {L167}
  (\bibinfo {year} {2004})}\BibitemShut {NoStop}%
\bibitem [{\citenamefont {Cravens}(2002)}]{cravens2002:sci}%
  \BibitemOpen
  \bibfield  {author} {\bibinfo {author} {\bibfnamefont {T.~E.}\ \bibnamefont
  {Cravens}},\ }\href {\doibase 10.1126/science.1070001} {\bibfield  {journal}
  {\bibinfo  {journal} {Science}\ }\textbf {\bibinfo {volume} {296}},\ \bibinfo
  {pages} {1042} (\bibinfo {year} {2002})}\BibitemShut {NoStop}%
\bibitem [{\citenamefont {Gutbier}(1957)}]{Gutbier1957:zfuna}%
  \BibitemOpen
  \bibfield  {author} {\bibinfo {author} {\bibfnamefont {H.}~\bibnamefont
  {Gutbier}},\ }\href {\doibase 10.1515/zna-1957-0606} {\bibfield  {journal}
  {\bibinfo  {journal} {Zeitschrift f{\"u}r Naturforschung A}\ }\textbf
  {\bibinfo {volume} {12}},\ \bibinfo {pages} {499} (\bibinfo {year}
  {1957})}\BibitemShut {NoStop}%
\bibitem [{\citenamefont {Stevenson}\ and\ \citenamefont
  {Schissler}(1958)}]{Stevenson1958:jcp}%
  \BibitemOpen
  \bibfield  {author} {\bibinfo {author} {\bibfnamefont {D.}~\bibnamefont
  {Stevenson}}\ and\ \bibinfo {author} {\bibfnamefont {D.}~\bibnamefont
  {Schissler}},\ }\href {\doibase 10.1063/1.1744476} {\bibfield  {journal}
  {\bibinfo  {journal} {J. Chem. Phys}\ }\textbf {\bibinfo {volume} {29}},\
  \bibinfo {pages} {282} (\bibinfo {year} {1958})}\BibitemShut {NoStop}%
\bibitem [{\citenamefont {Giese}, \citenamefont {F},\ and\ \citenamefont
  {Maier~II}(1963)}]{Giese1963:jcp}%
  \BibitemOpen
  \bibfield  {author} {\bibinfo {author} {\bibnamefont {Giese}}, \bibinfo
  {author} {\bibfnamefont {C.}~\bibnamefont {F}}, \ and\ \bibinfo {author}
  {\bibfnamefont {W.~B.}\ \bibnamefont {Maier~II}},\ }\href {\doibase
  /10.1063/1.1734318} {\bibfield  {journal} {\bibinfo  {journal} {J. Chem.
  Phys.}\ }\textbf {\bibinfo {volume} {39}},\ \bibinfo {pages} {739} (\bibinfo
  {year} {1963})}\BibitemShut {NoStop}%
\bibitem [{\citenamefont {Klein}\ and\ \citenamefont
  {Friedman}(1964)}]{Klein1964:jcp}%
  \BibitemOpen
  \bibfield  {author} {\bibinfo {author} {\bibfnamefont {F.~S.}\ \bibnamefont
  {Klein}}\ and\ \bibinfo {author} {\bibfnamefont {L.}~\bibnamefont
  {Friedman}},\ }\href {\doibase /10.1063/1.1726159} {\bibfield  {journal}
  {\bibinfo  {journal} {J. Chem. Phys.}\ }\textbf {\bibinfo {volume} {41}},\
  \bibinfo {pages} {1789} (\bibinfo {year} {1964})}\BibitemShut {NoStop}%
\bibitem [{\citenamefont {Aquilanti}\ \emph {et~al.}(1965)\citenamefont
  {Aquilanti}, \citenamefont {Galli}, \citenamefont {Giardini-Guidoni},\ and\
  \citenamefont {Volpi}}]{Aquilanti1965:jcp}%
  \BibitemOpen
  \bibfield  {author} {\bibinfo {author} {\bibfnamefont {V.}~\bibnamefont
  {Aquilanti}}, \bibinfo {author} {\bibfnamefont {A.}~\bibnamefont {Galli}},
  \bibinfo {author} {\bibfnamefont {A.}~\bibnamefont {Giardini-Guidoni}}, \
  and\ \bibinfo {author} {\bibfnamefont {G.}~\bibnamefont {Volpi}},\ }\href
  {\doibase /10.1063/1.1697061} {\bibfield  {journal} {\bibinfo  {journal} {J.
  Chem. Phys.}\ }\textbf {\bibinfo {volume} {43}},\ \bibinfo {pages} {1969}
  (\bibinfo {year} {1965})}\BibitemShut {NoStop}%
\bibitem [{\citenamefont {Henglein}, \citenamefont {Lacmann},\ and\
  \citenamefont {Knoll}(1965)}]{Henglein1965:jcp}%
  \BibitemOpen
  \bibfield  {author} {\bibinfo {author} {\bibfnamefont {A.}~\bibnamefont
  {Henglein}}, \bibinfo {author} {\bibfnamefont {K.}~\bibnamefont {Lacmann}}, \
  and\ \bibinfo {author} {\bibfnamefont {B.}~\bibnamefont {Knoll}},\ }\href
  {\doibase /10.1063/1.1696817} {\bibfield  {journal} {\bibinfo  {journal} {J.
  Chem. Phys.}\ }\textbf {\bibinfo {volume} {43}},\ \bibinfo {pages} {1048}
  (\bibinfo {year} {1965})}\BibitemShut {NoStop}%
\bibitem [{\citenamefont {Lacmann}\ and\ \citenamefont
  {Henglein}(1965)}]{Lacmann1965:bdbfupc}%
  \BibitemOpen
  \bibfield  {author} {\bibinfo {author} {\bibfnamefont {K.}~\bibnamefont
  {Lacmann}}\ and\ \bibinfo {author} {\bibfnamefont {A.}~\bibnamefont
  {Henglein}},\ }\href {\doibase /10.1002/bbpc.19650690405} {\bibfield
  {journal} {\bibinfo  {journal} {Ber. Bunsenges. Phys. Chem.}\ }\textbf
  {\bibinfo {volume} {69}},\ \bibinfo {pages} {286} (\bibinfo {year}
  {1965})}\BibitemShut {NoStop}%
\bibitem [{\citenamefont {Doverspike}, \citenamefont {Champion},\ and\
  \citenamefont {Bailey}(1966)}]{Doverspike1966:jcp}%
  \BibitemOpen
  \bibfield  {author} {\bibinfo {author} {\bibfnamefont {L.}~\bibnamefont
  {Doverspike}}, \bibinfo {author} {\bibfnamefont {R.}~\bibnamefont
  {Champion}}, \ and\ \bibinfo {author} {\bibfnamefont {T.}~\bibnamefont
  {Bailey}},\ }\href {\doibase /10.1063/1.1727517} {\bibfield  {journal}
  {\bibinfo  {journal} {J. Chem. Phys.}\ }\textbf {\bibinfo {volume} {45}},\
  \bibinfo {pages} {4385} (\bibinfo {year} {1966})}\BibitemShut {NoStop}%
\bibitem [{\citenamefont {Amme}\ and\ \citenamefont
  {McIlwain}(1966)}]{Amme1966:jcp}%
  \BibitemOpen
  \bibfield  {author} {\bibinfo {author} {\bibfnamefont {R.~C.}\ \bibnamefont
  {Amme}}\ and\ \bibinfo {author} {\bibfnamefont {J.~F.}\ \bibnamefont
  {McIlwain}},\ }\href {\doibase /10.1063/1.1727741} {\bibfield  {journal}
  {\bibinfo  {journal} {J. Chem. Phys.}\ }\textbf {\bibinfo {volume} {45}},\
  \bibinfo {pages} {1224} (\bibinfo {year} {1966})}\BibitemShut {NoStop}%
\bibitem [{\citenamefont {Ding}, \citenamefont {Lacmann},\ and\ \citenamefont
  {Henglein}(1967)}]{Ding1967:bdbfupc}%
  \BibitemOpen
  \bibfield  {author} {\bibinfo {author} {\bibfnamefont {A.}~\bibnamefont
  {Ding}}, \bibinfo {author} {\bibfnamefont {K.}~\bibnamefont {Lacmann}}, \
  and\ \bibinfo {author} {\bibfnamefont {A.}~\bibnamefont {Henglein}},\ }\href
  {\doibase 10.1002/bbpc.19670710610} {\bibfield  {journal} {\bibinfo
  {journal} {Ber. Bunsenges. Phys. Chem.}\ }\textbf {\bibinfo {volume} {71}},\
  \bibinfo {pages} {596} (\bibinfo {year} {1967})}\BibitemShut {NoStop}%
\bibitem [{\citenamefont {Fink}\ and\ \citenamefont
  {King~Jr}(1967)}]{Fink1967:jcp}%
  \BibitemOpen
  \bibfield  {author} {\bibinfo {author} {\bibfnamefont {R.~D.}\ \bibnamefont
  {Fink}}\ and\ \bibinfo {author} {\bibfnamefont {J.~S.}\ \bibnamefont
  {King~Jr}},\ }\href {\doibase 10.1063/1.1712182} {\bibfield  {journal}
  {\bibinfo  {journal} {J. Chem. Phys.}\ }\textbf {\bibinfo {volume} {47}},\
  \bibinfo {pages} {1857} (\bibinfo {year} {1967})}\BibitemShut {NoStop}%
\bibitem [{\citenamefont {Mahadevan}\ and\ \citenamefont
  {Magnuson}(1968)}]{Mahadevan1968:pr}%
  \BibitemOpen
  \bibfield  {author} {\bibinfo {author} {\bibfnamefont {P.}~\bibnamefont
  {Mahadevan}}\ and\ \bibinfo {author} {\bibfnamefont {G.}~\bibnamefont
  {Magnuson}},\ }\href {\doibase /10.1063/1.1712182} {\bibfield  {journal}
  {\bibinfo  {journal} {Phys. Rev.}\ }\textbf {\bibinfo {volume} {171}},\
  \bibinfo {pages} {103} (\bibinfo {year} {1968})}\BibitemShut {NoStop}%
\bibitem [{\citenamefont {Hyatt}\ and\ \citenamefont
  {Lacmann}(1968)}]{Hyatt1968:zn}%
  \BibitemOpen
  \bibfield  {author} {\bibinfo {author} {\bibfnamefont {D.}~\bibnamefont
  {Hyatt}}\ and\ \bibinfo {author} {\bibfnamefont {K.}~\bibnamefont
  {Lacmann}},\ }\href {\doibase /10.1515/zna-1968-1231} {\bibfield  {journal}
  {\bibinfo  {journal} {Z. Naturforsch.}\ }\textbf {\bibinfo {volume} {23}},\
  \bibinfo {pages} {2080} (\bibinfo {year} {1968})}\BibitemShut {NoStop}%
\bibitem [{\citenamefont {Chupka}\ and\ \citenamefont
  {Russell}(1968)}]{Chupka1968:jcp}%
  \BibitemOpen
  \bibfield  {author} {\bibinfo {author} {\bibfnamefont {W.}~\bibnamefont
  {Chupka}}\ and\ \bibinfo {author} {\bibfnamefont {M.~E.}\ \bibnamefont
  {Russell}},\ }\href {\doibase /10.1063/1.1670068} {\bibfield  {journal}
  {\bibinfo  {journal} {J. Chem. Phys.}\ }\textbf {\bibinfo {volume} {49}},\
  \bibinfo {pages} {5426} (\bibinfo {year} {1968})}\BibitemShut {NoStop}%
\bibitem [{\citenamefont {Bowers}\ and\ \citenamefont
  {Elleman}(1969)}]{Bowers1969:jcp}%
  \BibitemOpen
  \bibfield  {author} {\bibinfo {author} {\bibfnamefont {M.~T.}\ \bibnamefont
  {Bowers}}\ and\ \bibinfo {author} {\bibfnamefont {D.~D.}\ \bibnamefont
  {Elleman}},\ }\href {\doibase /10.1063/1.1671833} {\bibfield  {journal}
  {\bibinfo  {journal} {J. Chem. Phys.}\ }\textbf {\bibinfo {volume} {51}},\
  \bibinfo {pages} {4606} (\bibinfo {year} {1969})}\BibitemShut {NoStop}%
\bibitem [{\citenamefont {Adams}\ \emph {et~al.}(1970)\citenamefont {Adams},
  \citenamefont {Bohme}, \citenamefont {Dunkin},\ and\ \citenamefont
  {Fehsenfeld}}]{Adams1970:jcp}%
  \BibitemOpen
  \bibfield  {author} {\bibinfo {author} {\bibfnamefont {N.}~\bibnamefont
  {Adams}}, \bibinfo {author} {\bibfnamefont {D.}~\bibnamefont {Bohme}},
  \bibinfo {author} {\bibfnamefont {D.}~\bibnamefont {Dunkin}}, \ and\ \bibinfo
  {author} {\bibfnamefont {F.}~\bibnamefont {Fehsenfeld}},\ }\href {\doibase
  /10.1063/1.1673239} {\bibfield  {journal} {\bibinfo  {journal} {J. Chem.
  Phys.}\ }\textbf {\bibinfo {volume} {52}},\ \bibinfo {pages} {1951} (\bibinfo
  {year} {1970})}\BibitemShut {NoStop}%
\bibitem [{\citenamefont {Chiang}\ \emph {et~al.}(1970)\citenamefont {Chiang},
  \citenamefont {Gislason}, \citenamefont {Mahan}, \citenamefont {Tsao},\ and\
  \citenamefont {Werner}}]{Chiang1970:jcp}%
  \BibitemOpen
  \bibfield  {author} {\bibinfo {author} {\bibfnamefont {M.}~\bibnamefont
  {Chiang}}, \bibinfo {author} {\bibfnamefont {E.}~\bibnamefont {Gislason}},
  \bibinfo {author} {\bibfnamefont {B.}~\bibnamefont {Mahan}}, \bibinfo
  {author} {\bibfnamefont {C.}~\bibnamefont {Tsao}}, \ and\ \bibinfo {author}
  {\bibfnamefont {A.}~\bibnamefont {Werner}},\ }\href {\doibase
  /10.1063/1.1673361} {\bibfield  {journal} {\bibinfo  {journal} {J. Chem.
  Phys.}\ }\textbf {\bibinfo {volume} {52}},\ \bibinfo {pages} {2698} (\bibinfo
  {year} {1970})}\BibitemShut {NoStop}%
\bibitem [{\citenamefont {Ryan}\ and\ \citenamefont
  {Graham}(1973)}]{Ryan1973:jcp}%
  \BibitemOpen
  \bibfield  {author} {\bibinfo {author} {\bibfnamefont {K.}~\bibnamefont
  {Ryan}}\ and\ \bibinfo {author} {\bibfnamefont {I.}~\bibnamefont {Graham}},\
  }\href {\doibase /10.1016/0042-207x(75)91264-6} {\bibfield  {journal}
  {\bibinfo  {journal} {J. Chem. Phys.}\ }\textbf {\bibinfo {volume} {59}},\
  \bibinfo {pages} {4260} (\bibinfo {year} {1973})}\BibitemShut {NoStop}%
\bibitem [{\citenamefont {Teloy}\ and\ \citenamefont
  {Gerlich}(1974)}]{Teloy1974:cp}%
  \BibitemOpen
  \bibfield  {author} {\bibinfo {author} {\bibfnamefont {E.}~\bibnamefont
  {Teloy}}\ and\ \bibinfo {author} {\bibfnamefont {D.}~\bibnamefont
  {Gerlich}},\ }\href {\doibase 10.1016/0301-0104(74)85008-1} {\bibfield
  {journal} {\bibinfo  {journal} {Chem. Phys.}\ }\textbf {\bibinfo {volume}
  {4}},\ \bibinfo {pages} {417} (\bibinfo {year} {1974})}\BibitemShut {NoStop}%
\bibitem [{\citenamefont {Smith}, \citenamefont {Smith},\ and\ \citenamefont
  {Futrell}(1976)}]{Smith1976:ijmsip}%
  \BibitemOpen
  \bibfield  {author} {\bibinfo {author} {\bibfnamefont {R.~D.}\ \bibnamefont
  {Smith}}, \bibinfo {author} {\bibfnamefont {D.}~\bibnamefont {Smith}}, \ and\
  \bibinfo {author} {\bibfnamefont {J.}~\bibnamefont {Futrell}},\ }\href
  {\doibase /10.1016/0020-7381(76)80021-6} {\bibfield  {journal} {\bibinfo
  {journal} {IJMSIP}\ }\textbf {\bibinfo {volume} {19}},\ \bibinfo {pages}
  {395} (\bibinfo {year} {1976})}\BibitemShut {NoStop}%
\bibitem [{\citenamefont {Hodge~Jr}\ \emph {et~al.}(1977)\citenamefont
  {Hodge~Jr}, \citenamefont {Goldberger}, \citenamefont {Vedder},\ and\
  \citenamefont {Pollack}}]{HodgeJr1977:pra}%
  \BibitemOpen
  \bibfield  {author} {\bibinfo {author} {\bibfnamefont {W.}~\bibnamefont
  {Hodge~Jr}}, \bibinfo {author} {\bibfnamefont {A.}~\bibnamefont
  {Goldberger}}, \bibinfo {author} {\bibfnamefont {M.}~\bibnamefont {Vedder}},
  \ and\ \bibinfo {author} {\bibfnamefont {E.}~\bibnamefont {Pollack}},\ }\href
  {\doibase /10.1103/physreva.16.2360} {\bibfield  {journal} {\bibinfo
  {journal} {Phys. Rev. A}\ }\textbf {\bibinfo {volume} {16}},\ \bibinfo
  {pages} {2360} (\bibinfo {year} {1977})}\BibitemShut {NoStop}%
\bibitem [{\citenamefont {Hierl}, \citenamefont {Herman},\ and\ \citenamefont
  {Pacak}(1977)}]{Hierl1977:jcp}%
  \BibitemOpen
  \bibfield  {author} {\bibinfo {author} {\bibfnamefont {P.~M.}\ \bibnamefont
  {Hierl}}, \bibinfo {author} {\bibfnamefont {Z.}~\bibnamefont {Herman}}, \
  and\ \bibinfo {author} {\bibfnamefont {V.}~\bibnamefont {Pacak}},\ }\href
  {\doibase /10.1063/1.435181} {\bibfield  {journal} {\bibinfo  {journal} {J.
  Chem. Phys.}\ }\textbf {\bibinfo {volume} {67}},\ \bibinfo {pages} {2678}
  (\bibinfo {year} {1977})}\BibitemShut {NoStop}%
\bibitem [{\citenamefont {Lindinger}\ \emph {et~al.}(1977)\citenamefont
  {Lindinger}, \citenamefont {Alge}, \citenamefont {St{\"o}ri}, \citenamefont
  {Pahl},\ and\ \citenamefont {Varney}}]{Lindinger1977:jcp}%
  \BibitemOpen
  \bibfield  {author} {\bibinfo {author} {\bibfnamefont {W.}~\bibnamefont
  {Lindinger}}, \bibinfo {author} {\bibfnamefont {E.}~\bibnamefont {Alge}},
  \bibinfo {author} {\bibfnamefont {H.}~\bibnamefont {St{\"o}ri}}, \bibinfo
  {author} {\bibfnamefont {M.}~\bibnamefont {Pahl}}, \ and\ \bibinfo {author}
  {\bibfnamefont {R.}~\bibnamefont {Varney}},\ }\href {\doibase
  /10.1063/1.435346} {\bibfield  {journal} {\bibinfo  {journal} {J. Chem.
  Phys.}\ }\textbf {\bibinfo {volume} {67}},\ \bibinfo {pages} {3495} (\bibinfo
  {year} {1977})}\BibitemShut {NoStop}%
\bibitem [{\citenamefont {Rakshit}\ and\ \citenamefont
  {Warneck}(1980)}]{Rakshit1980:jcp}%
  \BibitemOpen
  \bibfield  {author} {\bibinfo {author} {\bibfnamefont {A.}~\bibnamefont
  {Rakshit}}\ and\ \bibinfo {author} {\bibfnamefont {P.}~\bibnamefont
  {Warneck}},\ }\href {\doibase /10.1063/1.440480} {\bibfield  {journal}
  {\bibinfo  {journal} {J. Chem. Phys.}\ }\textbf {\bibinfo {volume} {73}},\
  \bibinfo {pages} {2673} (\bibinfo {year} {1980})}\BibitemShut {NoStop}%
\bibitem [{\citenamefont {Tanaka}\ \emph {et~al.}(1981)\citenamefont {Tanaka},
  \citenamefont {Durup}, \citenamefont {Kato},\ and\ \citenamefont
  {Koyano}}]{Tanaka1981:jcp}%
  \BibitemOpen
  \bibfield  {author} {\bibinfo {author} {\bibfnamefont {K.}~\bibnamefont
  {Tanaka}}, \bibinfo {author} {\bibfnamefont {J.}~\bibnamefont {Durup}},
  \bibinfo {author} {\bibfnamefont {T.}~\bibnamefont {Kato}}, \ and\ \bibinfo
  {author} {\bibfnamefont {I.}~\bibnamefont {Koyano}},\ }\href {\doibase
  /10.1063/1.440919} {\bibfield  {journal} {\bibinfo  {journal} {J. Chem.
  Phys.}\ }\textbf {\bibinfo {volume} {74}},\ \bibinfo {pages} {5561} (\bibinfo
  {year} {1981})}\BibitemShut {NoStop}%
\bibitem [{\citenamefont {Dotan}\ and\ \citenamefont
  {Lindinger}(1982)}]{Dotan1982:jcp}%
  \BibitemOpen
  \bibfield  {author} {\bibinfo {author} {\bibfnamefont {I.}~\bibnamefont
  {Dotan}}\ and\ \bibinfo {author} {\bibfnamefont {W.}~\bibnamefont
  {Lindinger}},\ }\href {\doibase /10.1063/1.442843} {\bibfield  {journal}
  {\bibinfo  {journal} {J. Chem. Phys.}\ }\textbf {\bibinfo {volume} {76}},\
  \bibinfo {pages} {4972} (\bibinfo {year} {1982})}\BibitemShut {NoStop}%
\bibitem [{\citenamefont {Kemper}\ and\ \citenamefont
  {Bowers}(1983)}]{Kemper1983:i}%
  \BibitemOpen
  \bibfield  {author} {\bibinfo {author} {\bibfnamefont {P.~R.}\ \bibnamefont
  {Kemper}}\ and\ \bibinfo {author} {\bibfnamefont {M.~T.}\ \bibnamefont
  {Bowers}},\ }\href {\doibase /10.1016/0020-7381(83)85088-8} {\bibfield
  {journal} {\bibinfo  {journal} {IJMSIP}\ }\textbf {\bibinfo {volume} {52}},\
  \bibinfo {pages} {1} (\bibinfo {year} {1983})}\BibitemShut {NoStop}%
\bibitem [{\citenamefont {Kato}(1984)}]{Kato1984:jcp}%
  \BibitemOpen
  \bibfield  {author} {\bibinfo {author} {\bibfnamefont {T.}~\bibnamefont
  {Kato}},\ }\href {\doibase /10.1063/1.446711} {\bibfield  {journal} {\bibinfo
   {journal} {J. Chem. Phys.}\ }\textbf {\bibinfo {volume} {80}},\ \bibinfo
  {pages} {6105} (\bibinfo {year} {1984})}\BibitemShut {NoStop}%
\bibitem [{\citenamefont {Hamdan}, \citenamefont {Birkinshaw},\ and\
  \citenamefont {Twiddy}(1984)}]{Hamdan1984:ijmsip}%
  \BibitemOpen
  \bibfield  {author} {\bibinfo {author} {\bibfnamefont {M.}~\bibnamefont
  {Hamdan}}, \bibinfo {author} {\bibfnamefont {K.}~\bibnamefont {Birkinshaw}},
  \ and\ \bibinfo {author} {\bibfnamefont {N.}~\bibnamefont {Twiddy}},\ }\href
  {\doibase h/10.1016/0168-1176(84)87116-5} {\bibfield  {journal} {\bibinfo
  {journal} {Int. J. Mass Spectrom. Ion Processes}\ }\textbf {\bibinfo {volume}
  {62}},\ \bibinfo {pages} {297} (\bibinfo {year} {1984})}\BibitemShut
  {NoStop}%
\bibitem [{\citenamefont {Ervin}\ and\ \citenamefont
  {Armentrout}(1985)}]{Ervin1985:jcp}%
  \BibitemOpen
  \bibfield  {author} {\bibinfo {author} {\bibfnamefont {K.~M.}\ \bibnamefont
  {Ervin}}\ and\ \bibinfo {author} {\bibfnamefont {P.~B.}\ \bibnamefont
  {Armentrout}},\ }\href {\doibase /10.1063/1.449799} {\bibfield  {journal}
  {\bibinfo  {journal} {J. Chem. Phys.}\ }\textbf {\bibinfo {volume} {83}},\
  \bibinfo {pages} {166} (\bibinfo {year} {1985})}\BibitemShut {NoStop}%
\bibitem [{\citenamefont {Nakamura}, \citenamefont {Kobayashi},\ and\
  \citenamefont {Kaneko}(1986)}]{Nakamura1986:jpsj}%
  \BibitemOpen
  \bibfield  {author} {\bibinfo {author} {\bibfnamefont {T.}~\bibnamefont
  {Nakamura}}, \bibinfo {author} {\bibfnamefont {N.}~\bibnamefont {Kobayashi}},
  \ and\ \bibinfo {author} {\bibfnamefont {Y.}~\bibnamefont {Kaneko}},\ }\href
  {\doibase /10.1143/jpsj.55.3831} {\bibfield  {journal} {\bibinfo  {journal}
  {J. Phys. Soc. Jpn.}\ }\textbf {\bibinfo {volume} {55}},\ \bibinfo {pages}
  {3831} (\bibinfo {year} {1986})}\BibitemShut {NoStop}%
\bibitem [{\citenamefont {Henri}\ \emph {et~al.}(1988)\citenamefont {Henri},
  \citenamefont {Lavoll{\'e}e}, \citenamefont {Dutuit}, \citenamefont {Ozenne},
  \citenamefont {Guyon},\ and\ \citenamefont {Gislason}}]{Henri1988:jcp}%
  \BibitemOpen
  \bibfield  {author} {\bibinfo {author} {\bibfnamefont {G.}~\bibnamefont
  {Henri}}, \bibinfo {author} {\bibfnamefont {M.}~\bibnamefont {Lavoll{\'e}e}},
  \bibinfo {author} {\bibfnamefont {O.}~\bibnamefont {Dutuit}}, \bibinfo
  {author} {\bibfnamefont {J.}~\bibnamefont {Ozenne}}, \bibinfo {author}
  {\bibfnamefont {P.}~\bibnamefont {Guyon}}, \ and\ \bibinfo {author}
  {\bibfnamefont {E.}~\bibnamefont {Gislason}},\ }\href {\doibase
  /10.1063/1.454475} {\bibfield  {journal} {\bibinfo  {journal} {J. Chem.
  Phys.}\ }\textbf {\bibinfo {volume} {88}},\ \bibinfo {pages} {6381} (\bibinfo
  {year} {1988})}\BibitemShut {NoStop}%
\bibitem [{\citenamefont {Lindsay}\ and\ \citenamefont
  {Latimer}(1988)}]{Lindsay1988:jpbamop}%
  \BibitemOpen
  \bibfield  {author} {\bibinfo {author} {\bibfnamefont {B.}~\bibnamefont
  {Lindsay}}\ and\ \bibinfo {author} {\bibfnamefont {C.}~\bibnamefont
  {Latimer}},\ }\href {\doibase /10.1088/0953-4075/21/9/019} {\bibfield
  {journal} {\bibinfo  {journal} {J. Phys. B: At., Mol. Opt. Phys.}\ }\textbf
  {\bibinfo {volume} {21}},\ \bibinfo {pages} {1617} (\bibinfo {year}
  {1988})}\BibitemShut {NoStop}%
\bibitem [{\citenamefont {Liao}\ \emph {et~al.}(1990)\citenamefont {Liao},
  \citenamefont {Xu}, \citenamefont {Nourbakhsh}, \citenamefont {Flesch},
  \citenamefont {Baer},\ and\ \citenamefont {Ng}}]{Liao1990:jcp}%
  \BibitemOpen
  \bibfield  {author} {\bibinfo {author} {\bibfnamefont {C.-L.}\ \bibnamefont
  {Liao}}, \bibinfo {author} {\bibfnamefont {R.}~\bibnamefont {Xu}}, \bibinfo
  {author} {\bibfnamefont {S.}~\bibnamefont {Nourbakhsh}}, \bibinfo {author}
  {\bibfnamefont {G.}~\bibnamefont {Flesch}}, \bibinfo {author} {\bibfnamefont
  {M.}~\bibnamefont {Baer}}, \ and\ \bibinfo {author} {\bibfnamefont
  {C.}~\bibnamefont {Ng}},\ }\href {\doibase /10.1063/1.459671} {\bibfield
  {journal} {\bibinfo  {journal} {J. Chem. Phys.}\ }\textbf {\bibinfo {volume}
  {93}},\ \bibinfo {pages} {4832} (\bibinfo {year} {1990})}\BibitemShut
  {NoStop}%
\bibitem [{\citenamefont {Gislason}\ and\ \citenamefont
  {Parlant}(1991)}]{Gislason1991:jcp}%
  \BibitemOpen
  \bibfield  {author} {\bibinfo {author} {\bibfnamefont {E.~A.}\ \bibnamefont
  {Gislason}}\ and\ \bibinfo {author} {\bibfnamefont {G.}~\bibnamefont
  {Parlant}},\ }\href@noop {} {\bibfield  {journal} {\bibinfo  {journal} {J.
  Chem. Phys.}\ }\textbf {\bibinfo {volume} {94}},\ \bibinfo {pages} {6598}
  (\bibinfo {year} {1991})}\BibitemShut {NoStop}%
\bibitem [{\citenamefont {Hawley}\ and\ \citenamefont
  {Smith}(1992)}]{Hawley1992:jcp}%
  \BibitemOpen
  \bibfield  {author} {\bibinfo {author} {\bibfnamefont {M.}~\bibnamefont
  {Hawley}}\ and\ \bibinfo {author} {\bibfnamefont {M.~A.}\ \bibnamefont
  {Smith}},\ }\href {\doibase 10.1063/1.462394} {\bibfield  {journal} {\bibinfo
   {journal} {J. Chem. Phys.}\ }\textbf {\bibinfo {volume} {96}},\ \bibinfo
  {pages} {7440} (\bibinfo {year} {1992})}\BibitemShut {NoStop}%
\bibitem [{\citenamefont {Tosi}\ \emph {et~al.}(1993)\citenamefont {Tosi},
  \citenamefont {Dmitrijev}, \citenamefont {Soldo}, \citenamefont {Bassi},
  \citenamefont {Cappelletti}, \citenamefont {Pirani},\ and\ \citenamefont
  {Aquilanti}}]{Tosi1993:jcp}%
  \BibitemOpen
  \bibfield  {author} {\bibinfo {author} {\bibfnamefont {P.}~\bibnamefont
  {Tosi}}, \bibinfo {author} {\bibfnamefont {O.}~\bibnamefont {Dmitrijev}},
  \bibinfo {author} {\bibfnamefont {Y.}~\bibnamefont {Soldo}}, \bibinfo
  {author} {\bibfnamefont {D.}~\bibnamefont {Bassi}}, \bibinfo {author}
  {\bibfnamefont {D.}~\bibnamefont {Cappelletti}}, \bibinfo {author}
  {\bibfnamefont {F.}~\bibnamefont {Pirani}}, \ and\ \bibinfo {author}
  {\bibfnamefont {V.}~\bibnamefont {Aquilanti}},\ }\href {\doibase
  /10.1063/1.465312} {\bibfield  {journal} {\bibinfo  {journal} {J. Chem.
  Phys.}\ }\textbf {\bibinfo {volume} {99}},\ \bibinfo {pages} {985} (\bibinfo
  {year} {1993})}\BibitemShut {NoStop}%
\bibitem [{\citenamefont {Gerlich}(1993)}]{Gerlich1993:acp}%
  \BibitemOpen
  \bibfield  {author} {\bibinfo {author} {\bibfnamefont {D.}~\bibnamefont
  {Gerlich}},\ }\href {\doibase 10.1063/1.45274} {\bibfield  {journal}
  {\bibinfo  {journal} {AIP Conf. Proc.}\ }\textbf {\bibinfo {volume} {295}},\
  \bibinfo {pages} {607} (\bibinfo {year} {1993})}\BibitemShut {NoStop}%
\bibitem [{\citenamefont {Schweizer}, \citenamefont {Mark},\ and\ \citenamefont
  {Gerlich}(1994)}]{schweizer1994:ijm}%
  \BibitemOpen
  \bibfield  {author} {\bibinfo {author} {\bibfnamefont {M.}~\bibnamefont
  {Schweizer}}, \bibinfo {author} {\bibfnamefont {S.}~\bibnamefont {Mark}}, \
  and\ \bibinfo {author} {\bibfnamefont {D.}~\bibnamefont {Gerlich}},\ }\href
  {\doibase 10.1016/0168-1176(94)03974-7} {\bibfield  {journal} {\bibinfo
  {journal} {Int. J. Mass Spectrom. Ion Processes}\ }\textbf {\bibinfo {volume}
  {135}},\ \bibinfo {pages} {1} (\bibinfo {year} {1994})}\BibitemShut {NoStop}%
\bibitem [{\citenamefont {Sizun}\ \emph {et~al.}(1996)\citenamefont {Sizun},
  \citenamefont {Aguillon}, \citenamefont {Sidis}, \citenamefont {Zenevich},
  \citenamefont {Billing},\ and\ \citenamefont {Markovi{\'c}}}]{Sizun1996:cp}%
  \BibitemOpen
  \bibfield  {author} {\bibinfo {author} {\bibfnamefont {M.}~\bibnamefont
  {Sizun}}, \bibinfo {author} {\bibfnamefont {F.}~\bibnamefont {Aguillon}},
  \bibinfo {author} {\bibfnamefont {V.}~\bibnamefont {Sidis}}, \bibinfo
  {author} {\bibfnamefont {V.}~\bibnamefont {Zenevich}}, \bibinfo {author}
  {\bibfnamefont {G.}~\bibnamefont {Billing}}, \ and\ \bibinfo {author}
  {\bibfnamefont {N.}~\bibnamefont {Markovi{\'c}}},\ }\href {\doibase
  /10.1016/0301-0104(96)00101-2} {\bibfield  {journal} {\bibinfo  {journal}
  {Chem. Phys.}\ }\textbf {\bibinfo {volume} {209}},\ \bibinfo {pages} {327}
  (\bibinfo {year} {1996})}\BibitemShut {NoStop}%
\bibitem [{\citenamefont {Uiterwaal}\ \emph {et~al.}(1996)\citenamefont
  {Uiterwaal}, \citenamefont {Van Der~Weg}, \citenamefont {Van~Eck},
  \citenamefont {van Emmichoven},\ and\ \citenamefont
  {Niehaus}}]{Uiterwaal1996:cp}%
  \BibitemOpen
  \bibfield  {author} {\bibinfo {author} {\bibfnamefont {C.}~\bibnamefont
  {Uiterwaal}}, \bibinfo {author} {\bibfnamefont {J.}~\bibnamefont {Van
  Der~Weg}}, \bibinfo {author} {\bibfnamefont {J.}~\bibnamefont {Van~Eck}},
  \bibinfo {author} {\bibfnamefont {P.~Z.}\ \bibnamefont {van Emmichoven}}, \
  and\ \bibinfo {author} {\bibfnamefont {A.}~\bibnamefont {Niehaus}},\ }\href
  {\doibase /10.1016/0301-0104(96)00166-8} {\bibfield  {journal} {\bibinfo
  {journal} {Chem. Phys.}\ }\textbf {\bibinfo {volume} {209}},\ \bibinfo
  {pages} {195} (\bibinfo {year} {1996})}\BibitemShut {NoStop}%
\bibitem [{\citenamefont {Sizun}, \citenamefont {Song},\ and\ \citenamefont
  {Gislason}(2002)}]{Sizun2002:jcp}%
  \BibitemOpen
  \bibfield  {author} {\bibinfo {author} {\bibfnamefont {M.}~\bibnamefont
  {Sizun}}, \bibinfo {author} {\bibfnamefont {J.-B.}\ \bibnamefont {Song}}, \
  and\ \bibinfo {author} {\bibfnamefont {E.~A.}\ \bibnamefont {Gislason}},\
  }\href {\doibase /10.1063/1.1434989} {\bibfield  {journal} {\bibinfo
  {journal} {J. Chem. Phys.}\ }\textbf {\bibinfo {volume} {116}},\ \bibinfo
  {pages} {2888} (\bibinfo {year} {2002})}\BibitemShut {NoStop}%
\bibitem [{\citenamefont {Langevin}(1905)}]{Langevin1905:acp}%
  \BibitemOpen
  \bibfield  {author} {\bibinfo {author} {\bibfnamefont {M.}~\bibnamefont
  {Langevin}},\ }in\ \href@noop {} {\emph {\bibinfo {booktitle} {Ann. Chim.
  Phys.}}},\ Vol.~\bibinfo {volume} {5}\ (\bibinfo {year} {1905})\ pp.\
  \bibinfo {pages} {245--288}\BibitemShut {NoStop}%
\bibitem [{\citenamefont {Gioumousis}\ and\ \citenamefont
  {Stevenson}(1958)}]{Gioumousis1958:jcp}%
  \BibitemOpen
  \bibfield  {author} {\bibinfo {author} {\bibfnamefont {G.}~\bibnamefont
  {Gioumousis}}\ and\ \bibinfo {author} {\bibfnamefont {D.}~\bibnamefont
  {Stevenson}},\ }\href {\doibase 10.1063/1.1744477} {\bibfield  {journal}
  {\bibinfo  {journal} {J. Chem. Phys}\ }\textbf {\bibinfo {volume} {29}},\
  \bibinfo {pages} {294} (\bibinfo {year} {1958})}\BibitemShut {NoStop}%
\bibitem [{\citenamefont {Birkinshaw}\ \emph {et~al.}(1987)\citenamefont
  {Birkinshaw}, \citenamefont {Shukla}, \citenamefont {Howard},\ and\
  \citenamefont {Futrell}}]{birkinshaw1987:cp}%
  \BibitemOpen
  \bibfield  {author} {\bibinfo {author} {\bibfnamefont {K.}~\bibnamefont
  {Birkinshaw}}, \bibinfo {author} {\bibfnamefont {A.}~\bibnamefont {Shukla}},
  \bibinfo {author} {\bibfnamefont {S.~L.}\ \bibnamefont {Howard}}, \ and\
  \bibinfo {author} {\bibfnamefont {J.~H.}\ \bibnamefont {Futrell}},\ }\href
  {\doibase 10.1016/0301-0104(87)80227-6} {\bibfield  {journal} {\bibinfo
  {journal} {Chem. Phys.}\ }\textbf {\bibinfo {volume} {113}},\ \bibinfo
  {pages} {149} (\bibinfo {year} {1987})}\BibitemShut {NoStop}%
\bibitem [{\citenamefont {Candori}\ \emph {et~al.}(2001)\citenamefont
  {Candori}, \citenamefont {Cavalli}, \citenamefont {Pirani}, \citenamefont
  {Volpi}, \citenamefont {Cappelletti}, \citenamefont {Tosi},\ and\
  \citenamefont {Bassi}}]{candori2001:jcp}%
  \BibitemOpen
  \bibfield  {author} {\bibinfo {author} {\bibfnamefont {R.}~\bibnamefont
  {Candori}}, \bibinfo {author} {\bibfnamefont {S.}~\bibnamefont {Cavalli}},
  \bibinfo {author} {\bibfnamefont {F.}~\bibnamefont {Pirani}}, \bibinfo
  {author} {\bibfnamefont {A.}~\bibnamefont {Volpi}}, \bibinfo {author}
  {\bibfnamefont {D.}~\bibnamefont {Cappelletti}}, \bibinfo {author}
  {\bibfnamefont {P.}~\bibnamefont {Tosi}}, \ and\ \bibinfo {author}
  {\bibfnamefont {D.}~\bibnamefont {Bassi}},\ }\href {\doibase
  10.1063/1.1413980} {\bibfield  {journal} {\bibinfo  {journal} {J. Chem.
  Phys.}\ }\textbf {\bibinfo {volume} {115}},\ \bibinfo {pages} {8888}
  (\bibinfo {year} {2001})}\BibitemShut {NoStop}%
\bibitem [{\citenamefont {Trippel}\ \emph {et~al.}(2013)\citenamefont
  {Trippel}, \citenamefont {Stei}, \citenamefont {Cox},\ and\ \citenamefont
  {Wester}}]{Trippel2013:prl}%
  \BibitemOpen
  \bibfield  {author} {\bibinfo {author} {\bibfnamefont {S.}~\bibnamefont
  {Trippel}}, \bibinfo {author} {\bibfnamefont {M.}~\bibnamefont {Stei}},
  \bibinfo {author} {\bibfnamefont {J.~A.}\ \bibnamefont {Cox}}, \ and\
  \bibinfo {author} {\bibfnamefont {R.}~\bibnamefont {Wester}},\ }\href
  {\doibase /10.1103/physrevlett.110.163201} {\bibfield  {journal} {\bibinfo
  {journal} {Phys. Rev. Lett.}\ }\textbf {\bibinfo {volume} {110}},\ \bibinfo
  {pages} {163201} (\bibinfo {year} {2013})}\BibitemShut {NoStop}%
\bibitem [{\citenamefont {Latimer}\ and\ \citenamefont
  {Campbell}(1982)}]{Latimer1982:jpb}%
  \BibitemOpen
  \bibfield  {author} {\bibinfo {author} {\bibfnamefont {C.}~\bibnamefont
  {Latimer}}\ and\ \bibinfo {author} {\bibfnamefont {F.}~\bibnamefont
  {Campbell}},\ }\href {\doibase /10.1088/0022-3700/15/11/021} {\bibfield
  {journal} {\bibinfo  {journal} {J. Phys. B}\ }\textbf {\bibinfo {volume}
  {15}},\ \bibinfo {pages} {1765} (\bibinfo {year} {1982})}\BibitemShut
  {NoStop}%
\bibitem [{\citenamefont {Houle}\ \emph {et~al.}(1982)\citenamefont {Houle},
  \citenamefont {Anderson}, \citenamefont {Gerlich}, \citenamefont {Turner},\
  and\ \citenamefont {Lee}}]{Houle1982:jcp}%
  \BibitemOpen
  \bibfield  {author} {\bibinfo {author} {\bibfnamefont {F.}~\bibnamefont
  {Houle}}, \bibinfo {author} {\bibfnamefont {S.~L.}\ \bibnamefont {Anderson}},
  \bibinfo {author} {\bibfnamefont {D.}~\bibnamefont {Gerlich}}, \bibinfo
  {author} {\bibfnamefont {T.}~\bibnamefont {Turner}}, \ and\ \bibinfo {author}
  {\bibfnamefont {Y.-T.}\ \bibnamefont {Lee}},\ }\href@noop {} {\bibfield
  {journal} {\bibinfo  {journal} {J. Chem. Phys.}\ }\textbf {\bibinfo {volume}
  {77}},\ \bibinfo {pages} {748} (\bibinfo {year} {1982})}\BibitemShut
  {NoStop}%
\bibitem [{\citenamefont {Baer}\ \emph {et~al.}(1990)\citenamefont {Baer},
  \citenamefont {Liao}, \citenamefont {Xu}, \citenamefont {Flesch},
  \citenamefont {Nourbakhsh}, \citenamefont {Ng},\ and\ \citenamefont
  {Neuhauser}}]{Baer1990:jcp}%
  \BibitemOpen
  \bibfield  {author} {\bibinfo {author} {\bibfnamefont {M.}~\bibnamefont
  {Baer}}, \bibinfo {author} {\bibfnamefont {C.-L.}\ \bibnamefont {Liao}},
  \bibinfo {author} {\bibfnamefont {R.}~\bibnamefont {Xu}}, \bibinfo {author}
  {\bibfnamefont {G.}~\bibnamefont {Flesch}}, \bibinfo {author} {\bibfnamefont
  {S.}~\bibnamefont {Nourbakhsh}}, \bibinfo {author} {\bibfnamefont
  {C.}~\bibnamefont {Ng}}, \ and\ \bibinfo {author} {\bibfnamefont
  {D.}~\bibnamefont {Neuhauser}},\ }\href {\doibase /10.1063/1.458674}
  {\bibfield  {journal} {\bibinfo  {journal} {J. Chem. Phys.}\ }\textbf
  {\bibinfo {volume} {93}},\ \bibinfo {pages} {4845} (\bibinfo {year}
  {1990})}\BibitemShut {NoStop}%
\bibitem [{\citenamefont {Hu}\ \emph {et~al.}(2013)\citenamefont {Hu},
  \citenamefont {Xu}, \citenamefont {Liu}, \citenamefont {Tan},\ and\
  \citenamefont {Li}}]{Hu2013:jcp}%
  \BibitemOpen
  \bibfield  {author} {\bibinfo {author} {\bibfnamefont {M.}~\bibnamefont
  {Hu}}, \bibinfo {author} {\bibfnamefont {W.}~\bibnamefont {Xu}}, \bibinfo
  {author} {\bibfnamefont {X.}~\bibnamefont {Liu}}, \bibinfo {author}
  {\bibfnamefont {R.}~\bibnamefont {Tan}}, \ and\ \bibinfo {author}
  {\bibfnamefont {H.}~\bibnamefont {Li}},\ }\href {\doibase /10.1063/1.4803116}
  {\bibfield  {journal} {\bibinfo  {journal} {J. Chem. Phys.}\ }\textbf
  {\bibinfo {volume} {138}},\ \bibinfo {pages} {174305} (\bibinfo {year}
  {2013})}\BibitemShut {NoStop}%
\bibitem [{\citenamefont {Stei}\ \emph {et~al.}(2016)\citenamefont {Stei},
  \citenamefont {Carrascosa}, \citenamefont {Kainz}, \citenamefont {Kelkar},
  \citenamefont {Meyer}, \citenamefont {Szabo}, \citenamefont {Czako},\ and\
  \citenamefont {Wester}}]{stei2016:natc}%
  \BibitemOpen
  \bibfield  {author} {\bibinfo {author} {\bibfnamefont {M.}~\bibnamefont
  {Stei}}, \bibinfo {author} {\bibfnamefont {E.}~\bibnamefont {Carrascosa}},
  \bibinfo {author} {\bibfnamefont {M.~A.}\ \bibnamefont {Kainz}}, \bibinfo
  {author} {\bibfnamefont {A.~K.}\ \bibnamefont {Kelkar}}, \bibinfo {author}
  {\bibfnamefont {J.}~\bibnamefont {Meyer}}, \bibinfo {author} {\bibfnamefont
  {I.}~\bibnamefont {Szabo}}, \bibinfo {author} {\bibfnamefont
  {G.}~\bibnamefont {Czako}}, \ and\ \bibinfo {author} {\bibfnamefont
  {R.}~\bibnamefont {Wester}},\ }\href {\doibase /10.1038/nchem.2400}
  {\bibfield  {journal} {\bibinfo  {journal} {Nature Chemistry}\ }\textbf
  {\bibinfo {volume} {8}},\ \bibinfo {pages} {151} (\bibinfo {year}
  {2016})}\BibitemShut {NoStop}%
\bibitem [{\citenamefont {Stei}\ \emph {et~al.}(2013)\citenamefont {Stei},
  \citenamefont {von Vangerow}, \citenamefont {Otto}, \citenamefont {Kelkar},
  \citenamefont {Carrascosa}, \citenamefont {Best},\ and\ \citenamefont
  {Wester}}]{stei2013:jcp}%
  \BibitemOpen
  \bibfield  {author} {\bibinfo {author} {\bibfnamefont {M.}~\bibnamefont
  {Stei}}, \bibinfo {author} {\bibfnamefont {J.}~\bibnamefont {von Vangerow}},
  \bibinfo {author} {\bibfnamefont {R.}~\bibnamefont {Otto}}, \bibinfo {author}
  {\bibfnamefont {A.~H.}\ \bibnamefont {Kelkar}}, \bibinfo {author}
  {\bibfnamefont {E.}~\bibnamefont {Carrascosa}}, \bibinfo {author}
  {\bibfnamefont {T.}~\bibnamefont {Best}}, \ and\ \bibinfo {author}
  {\bibfnamefont {R.}~\bibnamefont {Wester}},\ }\href@noop {} {\bibfield
  {journal} {\bibinfo  {journal} {J. Chem. Phys.}\ }\textbf {\bibinfo {volume}
  {138}},\ \bibinfo {pages} {214201} (\bibinfo {year} {2013})}\BibitemShut
  {NoStop}%
\bibitem [{\citenamefont {Wester}(2014)}]{Wester2014:pccp}%
  \BibitemOpen
  \bibfield  {author} {\bibinfo {author} {\bibfnamefont {R.}~\bibnamefont
  {Wester}},\ }\href {\doibase /10.1039/c3cp53405g} {\bibfield  {journal}
  {\bibinfo  {journal} {Phys. Chem. Chem. Phys.}\ }\textbf {\bibinfo {volume}
  {16}},\ \bibinfo {pages} {396} (\bibinfo {year} {2014})}\BibitemShut
  {NoStop}%
\bibitem [{\citenamefont {Ramos}\ \emph {et~al.}(2009)\citenamefont {Ramos},
  \citenamefont {Tejeda}, \citenamefont {Fern{\'a}ndez},\ and\ \citenamefont
  {Montero}}]{Ramos2009:jpca}%
  \BibitemOpen
  \bibfield  {author} {\bibinfo {author} {\bibfnamefont {A.}~\bibnamefont
  {Ramos}}, \bibinfo {author} {\bibfnamefont {G.}~\bibnamefont {Tejeda}},
  \bibinfo {author} {\bibfnamefont {J.}~\bibnamefont {Fern{\'a}ndez}}, \ and\
  \bibinfo {author} {\bibfnamefont {S.}~\bibnamefont {Montero}},\ }\href
  {\doibase /10.1021/jp901700c} {\bibfield  {journal} {\bibinfo  {journal} {J.
  Phys. Chem. A}\ }\textbf {\bibinfo {volume} {113}},\ \bibinfo {pages} {8506}
  (\bibinfo {year} {2009})}\BibitemShut {NoStop}%
\bibitem [{\citenamefont {Hernandez-Vera}\ \emph {et~al.}(2017)\citenamefont
  {Hernandez-Vera}, \citenamefont {Gianturco}, \citenamefont {Wester},
  \citenamefont {da~Silva~Jr}, \citenamefont {Dulieu},\ and\ \citenamefont
  {Schiller}}]{hernandez2017:jcp}%
  \BibitemOpen
  \bibfield  {author} {\bibinfo {author} {\bibfnamefont {M.}~\bibnamefont
  {Hernandez-Vera}}, \bibinfo {author} {\bibfnamefont {F.~A.}\ \bibnamefont
  {Gianturco}}, \bibinfo {author} {\bibfnamefont {R.}~\bibnamefont {Wester}},
  \bibinfo {author} {\bibfnamefont {H.}~\bibnamefont {da~Silva~Jr}}, \bibinfo
  {author} {\bibfnamefont {O.}~\bibnamefont {Dulieu}}, \ and\ \bibinfo {author}
  {\bibfnamefont {S.}~\bibnamefont {Schiller}},\ }\href@noop {} {\bibfield
  {journal} {\bibinfo  {journal} {J. Chem. Phys.}\ }\textbf {\bibinfo {volume}
  {146}},\ \bibinfo {pages} {124310} (\bibinfo {year} {2017})}\BibitemShut
  {NoStop}%
\end{thebibliography}
\end{document}